\RequirePackage{ifpdf}
\documentclass[hyper]{JHEP3}

\pdfoutput=1
\usepackage{epsfig}
\usepackage{amsmath}
\usepackage{inputenc}
\usepackage{graphicx}
\usepackage{subfigure}
\usepackage{snapshot}
\usepackage{cite}
\usepackage{slashed}
\def\beq{\begin{equation}}
\def\eeq{\end{equation}}
\def\beeq{\begin{eqnarray}}
\def\eeeq{\end{eqnarray}}

\def\nn{\nonumber}

\def\vec#1{\mbox{\boldmath $#1$}}

\def\bsig{\bar\sigma}
\def\spL#1{\langle #1\rangle}
\def\spR#1{[#1]}
\def\spM#1{\langle #1]}
\def\beps{\bar\varepsilon}
\def\gp2{g^{\prime 2}}

\title{Probing the Colour Structure of the Top Quark Forward-Backward Asymmetry}

\author{Ben Gripaios$^a$, Andreas Papaefstathiou$^b$ and Bryan Webber$^a$\\
   $^a$Cavendish Laboratory, J.J.\ Thomson Avenue, Cambridge, UK\\
   $^b$Institut f\"ur Theoretische Physik, Universit\"at Z\"urich, Switzerland\\
        E-mail: \email{gripaios@hep.phy.cam.ac.uk}, \email{andreasp@physik.uzh.ch}, \email{webber@hep.phy.cam.ac.uk}
        }

\preprint{Cavendish-HEP-13/07\\
ZU-TH 19/13
}

\abstract{We point out that QCD coherence effects can help to identify
the colour structure of possible new physics contributions to the anomalously large forward-backward
asymmetry in top quark pair production.  New physics models that 
yield the same inclusive asymmetry make different predictions for its
dependence on the transverse momentum of the pair, if they have
different colour structures.  From both a fixed-order effective field
theory approach and Monte Carlo studies of specific models, we find
that an $s$-channel octet structure is preferred.
}


\begin{document} 

\section{Introduction}
\label{sec:intro}
The surprisingly large forward--backward asymmetry observed in
the production of top quark pairs at the
Tevatron~\cite{Arbazov:2007qb,Aaltonen:2008hc,Aaltonen:2011kc,Abazov:2011rq}
has given rise to many attempted explanations in terms of physics
beyond the Standard Model (BSM): for recent reviews
see for
example~\cite{Kamenik:2011wt,Westhoff:2011tq,AguilarSaavedra:2012ma}.
New physics models seek to account for an asymmetry that rises with
the top pair invariant mass and is about twice as large as the current
best evaluations of the Standard Model
prediction~\cite{Hollik:2011ps,Kuhn:2011ri,Bernreuther:2012sx}.
While the experimental and theoretical uncertainties are large enough
that this discrepancy may eventually be resolved without new
physics\footnote{Recent analyses of dileptonic top
  decays~\cite{Aaltonen:2013vaf,Abazov:2013wxa} find lepton
  asymmetries less than two standard deviations from the Standard
  Model predictions.},
it is important to constrain the BSM models with all possible relevant
information.  This has prompted us to consider the implications of the
dependence of the asymmetry on the transverse momentum of the top
pair, as reported by the CDF Collaboration~\cite{Aaltonen:2012it}.

The observed asymmetry is positive at low transverse momentum but
falls and becomes negative at higher values.  The leading-order QCD
prediction has a similar behaviour but lies below the data.  Again, it
could be the case that further data and more complete Standard Model
calculations would resolve this discrepancy.  However, if new physics
is invoked, then it should it explain the transverse momentum
dependence of the asymmetry as well as its invariant mass
dependence.\footnote{In recent work~\cite{Hoeche:2013mua} it was found that 
  combining $t\bar t+0$ and 1-jet QCD NLO matrix elements with parton
  showers can give reasonable agreement with the transverse momentum
  dependence but not the invariant mass dependence.}

The tendency of the QCD contribution to the asymmetry to decrease with
increasing transverse momentum of the top pair has a simple
explanation in terms of QCD coherence~\cite{Dokshitzer:1987nm},
as was pointed out in~\cite{Kuhn:2011ri,Skands:2012mm}.  The
contributing process $q\bar q\to t\bar t$ does not have an asymmetry
at lowest order, but it has a colour structure that produces an
asymmetry in higher orders.  The $s$-channel gluon exchange means that
colour flows predominantly from the incoming quark to the outgoing
top, and anticolour from the incoming antiquark to the outgoing
antitop.  Thus there is more violent acceleration of the colour and
anticolour sources in backward top production that in forward, leading to
more QCD radiation in backward production, as depicted in
Fig.~\ref{fig:ttbar_coh}.  The emission of more radiation implies
a larger recoil of the top pair, so that higher transverse momentum of
the pair is correlated with a more backward top, and an asymmetry
that decreases with increasing transverse momentum is generated.  In
contrast, an $s$-channel colour-singlet mechanism would imply no
correlation between the amount of radiation and the production angle,
as in Fig.~\ref{fig:ttbar_sing}, and therefore no correlation between
the asymmetry and the transverse momentum of the pair.

 \FIGURE{
  \centering\centerline{
  \includegraphics[scale=0.5]{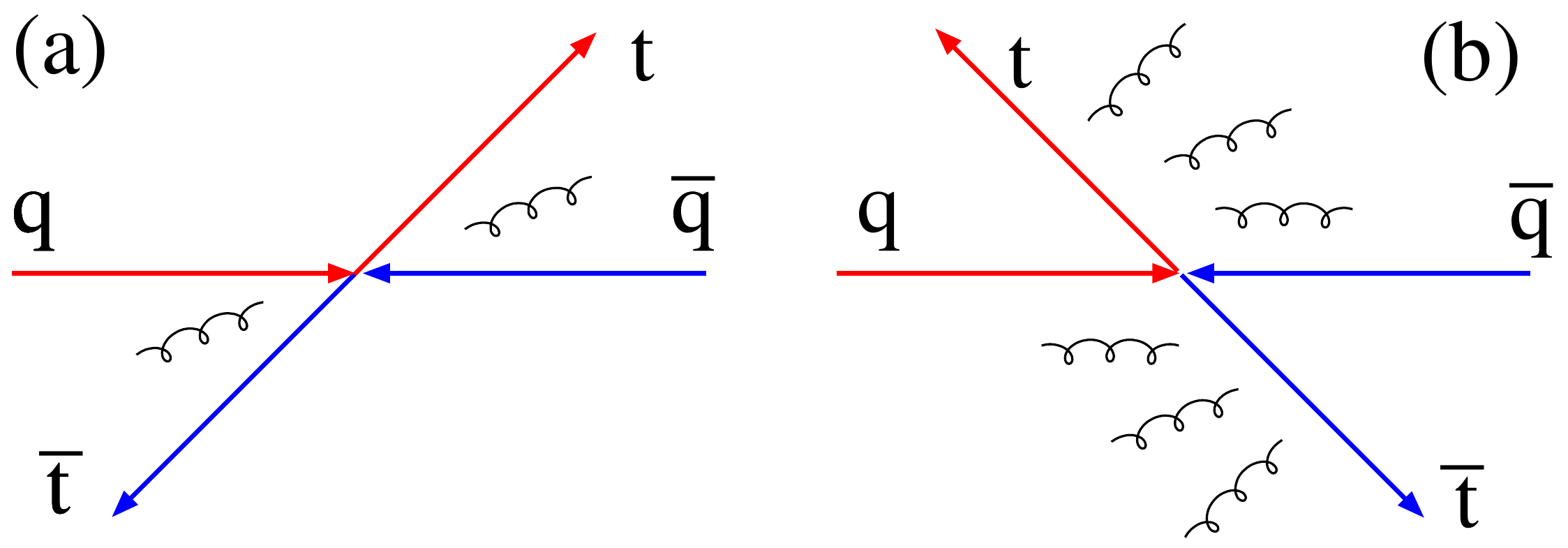}
  \caption{\label{fig:ttbar_coh}%
QCD radiation in $q\bar q\to t\bar t$ with an $s$-channel colour octet mechanism.  There is less radiation when the top quark goes forwards (a) and more when it goes backwards (b).}}
}

\FIGURE{
  \centering\centerline{
  \includegraphics[scale=0.5]{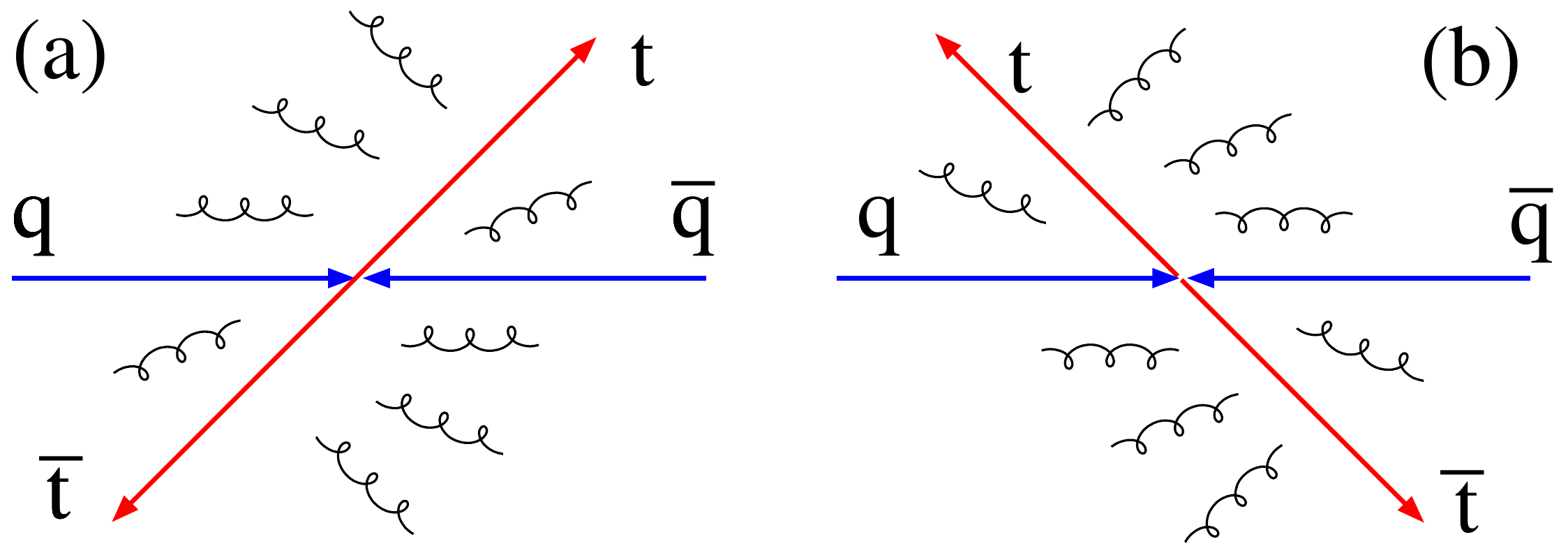}
  \caption{\label{fig:ttbar_sing}%
QCD radiation in $q\bar q\to t\bar t$ with an $s$-channel colour singlet mechanism.  The amount of radiation is the same in forward (a) and backward (b) production.}}
}

Now if we apply the same logic to a new physics process that
produces a positive asymmetry at lowest order, then we expect this asymmetry to
be modified at non-zero transverse momentum by recoil effects if the
process has an $s$-channel colour octet component. Backward top production
will still be correlated with greater recoil momenta, so the asymmetry
will be decreased relative to the lowest order.  On the other hand, an
$s$-channel singlet mechanism will not lead to any such change in the
asymmetry relative to the lowest order.\footnote{We should emphasise that we
are referring here to the effects of real gluon emission, at strictly
non-zero transverse momentum.  At each perturbative order, there is a
divergent virtual contribution at zero transverse momentum, and so the
average asymmetry at non-zero values is not directly related to the
inclusive asymmetry.}

To illustrate these ideas more quantitatively, we introduce in the
following Section an effective four-fermion interaction, representing
some mechanism beyond the Standard Model that can give rise to a
forward-backward asymmetry at the Born level.   We then examine the
transverse momentum dependent asymmetry that would arise from
gluon emission in such an interaction, contrasting the cases of
$s$-channel octet and singlet colour structures and including
interference with the QCD amplitude in the former. In each case we
search for the values of the four-fermion couplings that give the best
fits to the CDF data on the invariant mass and transverse momentum
dependences of the asymmetry.  In Section~\ref{sec:MC} we present
results from the {\tt HERWIG++} event generator for two BSM
models that illustrate the same points, and in Section~\ref{sec:conc}
we summarize our conclusions.  In the Appendix we present an analysis
neglecting the top quark mass, which has the advantages that the relevant
amplitudes can be given in a compact form and that the essential
qualitative features of the predictions remain valid.
 
\section{Four-fermion interaction model}
\subsection{Operator basis}
We begin by asking which four-fermion operators generate an asymmetry
at Born level. There are, {\em a priori}, infinitely many operators from
which to choose, since
operators of a given dimension in an effective field theory
form a complex vector space. We are only interested in Lagrangian
operators, which must be
Hermitian and invariant under Lorentz and gauge transformations, and which span
a real subspace. This subspace can be characterized by identifying a
convenient basis. 

Now, it is {\em not} sufficient to simply ask which of these basis elements generate
an asymmetry at Born level: While the Lagrangian operators have the
structure of a vector space, the asymmetries they result in (which are
obtained from matrix elements {\em squared}) do not. Therefore, to
fully answer our question, we
must consider the asymmetry that results from an arbitrary linear
combination of the basis elements.

With this {\em caveat} acknowledged, we proceed to
construct a basis of Lorentz and
gauge-invariant (under $SU(3)\times U(1)$ symmetries, corresponding to QCD and
electromagnetism\footnote{Note that other authors
  \cite{AguilarSaavedra:2010zi,Degrande:2010kt} restrict to operators
  symmetric under the full $SU(3)\times SU(2) \times U(1)$ group of
  the Standard Model.}) dimension six operators
involving a light quark-antiquark pair $q,\bar q$, and a heavy quark antiquark
pair $q',\bar q'$, leaving aside the issue of hermiticity for the time
being.

There are two possible $SU(3)$ colour structures, which correspond to
decomposing each quark-antiquark pair into either a singlet or an octet
representation. (Equivalently, each colour index of a given quark can
be contracted with one of two anti-quarks, giving two
linearly-independent operators.) It is these two possibilities that we hope to be able
to discriminate using the distribution of the asymmetry in transverse momentum.

The analysis of Lorentz structures is simplified 
by the presence of Fierz identities, which allow us to interchange the two fermions
in any 4-fermion operator.\footnote{Since the two fermion bi-linears in such an
operator commute, we can equivalently interchange the two
anti-fermions.} To be explicit, the most general form of the Fierz
identity can be obtained 
by taking any basis $\{ \Gamma^A \}$ for the vector space of complex, $4
\times 4$ matrices and writing the
completeness relation on this space in the form $\delta_{ij} \delta_{kl} = \sum_A
(\Gamma_A)_{il} (\Gamma^A)_{kj}$, where $\{ \Gamma_A \}$ is the basis dual
to $\{ \Gamma^A \}$. From this one obtains the Fierz identity $\Gamma^A_{ij}
\Gamma^B_{kl} = \sum_{C,D}
\mathrm{tr} (\Gamma^A \Gamma_D \Gamma^B \Gamma_C) \Gamma^C_{il}
\Gamma^D_{kj} $.
Using this identity, we can write all 4-fermion operators in the
ordered form
\beq
(\bar qM q)  (\bar q'M' q'),
\eeq
where $M,M'$ are arbitrary complex, $4\times 4$ matrices. (Proof: since
bi-linears commute, we can, without loss of generality, write the one involving $\bar q$ first; if this
bi-linear involves $q'$, use a Fierz identity to exchange it with
$q$.)

Next, we choose a particular basis
$\{\Gamma^A \} = \{ P_L, P_R, \gamma^\mu P_L, \gamma^\mu P_R, \sigma^{\mu
  \nu}\}$  for complex, $4\times 4$ matrices.  Out of these, we can
form 10 Lorentz invariant four-fermion operators, namely those with
Lorentz tensor combinations of the form $\{
P_{L,R}\otimes P_{L,R}, \gamma^\mu P_{L,R} \otimes \gamma_\mu P_{L,R}, \sigma^{\mu
  \nu}\otimes\sigma^{\mu
  \nu}, \epsilon_{\mu \nu \sigma \rho} \sigma^{\mu
  \nu}\otimes\sigma^{\sigma
  \rho}\}$.

Finally, we must enforce the restriction of hermiticity. One can
easily check that, amongst the four scalar combinations in the list, only $LL + RR$ and $LR + RL$ are Hermitian,
reducing the dimension of the subspace of Lagrangian four-fermion
operators to eight (or rather sixteen, once we include
the two possible colour structures).

Having found a basis for the vector space of Lagrangian operators, we
now ask which linear combinations of them can generate an asymmetry at Born level. This
asymmetry can either arise directly from the BSM operator, or via
interference with QCD.
As
we shall see below,
operators built out of the (Lorentz) vector combinations $\gamma^\mu P_{L,R}
\otimes \gamma_\mu
P_{L,R}$ generate an asymmetry on their own (provided that the left and
right couplings do not coincide), and in interfering with QCD if they
are colour octets. These operators, moreover, are the ones obtained by
integrating out the most interesting new
physics candidates, such as an axigluon or a $Z'$. 
Operators built purely out of (Lorentz) scalar or tensor
combinations do not interfere with QCD and do not generate an
asymmetry on their own. Indeed, the only other combination of our
basis elements that gives rise to an asymmetry involves the interference
between scalar
and tensor combinations. We do not
pursue this possibility further here.

\subsection{Born level cross-section}
\label{sec:born}
We thus consider a model in which a forward-backward asymmetry in $q\bar
q\to q'\bar q'$  is produced by an effective four-fermion interaction of the form
\beq\label{eq:4fermi}
\bar q\gamma^\mu(g_L P_L+g_R P_R) q\,
\bar q'\gamma_\mu(g'_L P_L+g'_R P_R) q'\,,
\eeq
where $q$ is massless and $q'$ has mass $m$.
Note that $g_{L,R}$ and $g'_{L,R}$ have dimension
1/mass.\footnote{Since the Lagrangian contains bi-linear
  functions of $g_{L,R}$ and $g'_{L,R}$, the physics will be invariant
  under a simultaneous change of sign of all of them. Similarly, since
  our observables are parity symmetric, cross-section formul\ae\ 
  will be invariant under $g^{(\prime )}_L \leftrightarrow g^{(\prime
    )}_R$.}

We treat all momenta as outgoing, i.e.
\beq
q(-p_1) +\bar q(-p_2)\to q'(p_3)+ \bar q'(p_4)\,,
\eeq
and define $s_{ij}=(p_i+p_j)^2$.  At Born level we have
$s_{12}=s_{34}=s$, $s_{13}=s_{24}=t$ and $s_{23}=s_{14}=u$. Then the
associated subprocess cross section is
\beeq\label{eq:dsdt}
\frac{d\sigma_{\rm BSM}}{dt} &=& \frac 1{16\pi s^2}
\bigl[(g_L^2\gp2_L+g_R^2\gp2_R)(u-m^2)^2+(g_L^2\gp2_R+g_R^2\gp2_L)(t-m^2)^2\nn\\
&&+2(g_L^2+g_R^2)g'_Lg'_R m^2 s\bigr]\,.
\eeeq
The corresponding QCD cross section has $g_{L,R}=g'_{L,R}=g_s$, a propagator
factor of $1/s^2$ and a colour factor of $C_F/2N$. Thus
\beq
\frac{d\sigma_{\rm QCD}}{dt} = \frac {g_s^4}{16\pi s^4}\frac{C_F}N
\left[(u-m^2)^2+(t-m^2)^2+2m^2 s\right]\,.
\eeq
If, on the one hand, the BSM interaction is $s$-channel colour-singlet, the two
contributions do not interfere and the forward and backward cross sections are
\beeq
&&\sigma^{\rm sing}_{\rm F,B} =\frac 1{96\pi s^2}\sqrt{1-\frac{4m^2}s}
\Bigl\{\left[(g_L^2+g_R^2)(\gp2_L+\gp2_R)s^2+2C_F
  g_s^4/N\right](s-m^2)\nn\\
&&+6\left[(g_L^2+g_R^2)g'_Lg'_R s^2 +C_F g_s^4/N\right]m^2\Bigr\}
\pm\frac 1{128\pi}(g_L^2-g_R^2)(\gp2_L-\gp2_R)(s-4m^2)\,.
\eeeq
If, on the other hand, the BSM contribution is $s$-channel colour-octet,
then the colour factors are the same and the two  contributions interfere:
\beeq
\frac{d\sigma^{\rm oct}}{dt} = \frac{C_F}{32\pi s^4N}&\Bigl\{&
\left[(g_Lg'_Ls+g_s^2)^2+(g_Rg'_Rs+g_s^2)^2\right](u-m^2)^2\nn\\
&+&
\left[(g_Lg'_Rs+g_s^2)^2+(g_Rg'_Ls+g_s^2)^2\right](t-m^2)^2 \nn\\
&+&2\left[(g_L^2+g_R^2)g'_Lg'_R s^2+(g_L+g_R)(g'_L+g'_R)g_s^2 s
+2g_s^4\right]m^2 s\;\Bigr\}\,,
\eeeq
giving
\beeq
\sigma^{\rm oct}_{\rm F,B} &=&\frac{C_F}{192\pi s^2N}\sqrt{1-\frac{4m^2}s}
\Bigl\{\left[(g_L^2+g_R^2)(\gp2_L+\gp2_R)s^2
+2(g_L+g_R)(g'_L+g'_R)g_s^2s +4g_s^4\right](s-m^2)\nn\\
&&+6\left[(g_L^2+g_R^2)g'_Lg'_R s^2+(g_L+g_R)(g'_L+g'_R)g_s^2 s
+2g_s^4\right]m^2 \Bigr\}\nn\\
&&\pm\frac{C_F}{128\pi sN}\left[(g_L^2-g_R^2)(\gp2_L-\gp2_R)s
+2(g_L-g_R)(g'_L-g'_R)g_s^2\right](s-4m^2)\,.
\eeeq

\subsection{One gluon emission}
\label{sec:1gluon}
The squared matrix elements for one-gluon emission in massive quark
pair production are rather cumbersome, and we do not include them
here.  In the limit of negligible quark mass, helicity amplitude
methods render the calculation much simpler, and it turns out that the
important features of the effects we wish to study are manifest, so we
discuss the massless case in some detail in the Appendix.  To obtain the
massive case results presented here,\footnote{Computation of the amplitudes in the massive case, both for real
  emission and the accompanying one-loop virtual diagrams, is reported in
  \cite{Shao:2011wa}, but detailed results are not given.} we revert to the old-fashioned
approach {\em \`{a} la} Feynman, based on squared matrix elements
rather than amplitudes, and averaged/summed over initial/final state
helicities.\footnote{Doing so also provides an independent check of
  our   results for $m =0$.} We use {\tt FEYNCALC}~\cite{Mertig:1990an} to
perform the more tedious Dirac algebra.

\subsection{Colour structure}

In the case of the four-fermion BSM interaction (\ref{eq:4fermi}),
for each combination of external helicities we can
associate an amplitude $A^{(i)}$ with emission of a gluon from
external line $i=1,\ldots,4$.  This is not a gauge-invariant
procedure, but the full matrix element-squared can be represented
as a sum of gauge-invariant combinations
\beq\label{eq:colour}
{\cal M} = \sum_{i<j}C_{ij}|A^{(i)}-A^{(j)}|^2\,,
\eeq
where the coefficients $C_{ij}$, averaged over initial and summed over
final colours, are given in Table~\ref{tab:colour}.  The QCD
expressions, and the QCD-BSM interference,  can also be written in this form.
 The gauge-invariant combinations are given in the massless
 case by eqs.~(\ref{eq:bsm1}), (\ref{eq:qcd1}) and (\ref{eq:int1}) in
 the Appendix. We have checked that for pure QCD, our results for the massive
case reproduce the expression given in \cite{Ellis:1986ef}.

We see from Table~\ref{tab:colour} that in the singlet case only the
(12) and (34) terms contribute, and therefore we have the
situation in Fig.~\ref{fig:ttbar_sing}, where the amount of QCD
radiation is the same in forward and backward production.
In the case of an octet interaction, on the other hand, the (13) and
(24) terms are dominant, and there is more radiation in backward
production, which correlates a higher transverse momentum of the
heavy quark pair with a more negative asymmetry than that at Born level.

\begin{table}
  \begin{center}    
    \begin{tabular}{|c|c|c|}
      \hline
      $ij$ & Singlet & Octet \\
      \hline
 12, 34 & $C_F$ & $-C_F/4N^2$ \\
 13, 24 & 0 & $(C_F/2N)(C_F-1/2N)$ \\
 14, 23 & 0 & $C_F/2N^2$ \\
 \hline
    \end{tabular}
  \end{center}
  \caption{
\label{tab:colour}Colour factors $C_{ij}$ for one gluon emission, for
$s$-channel singlet or octet.
}
\end{table}

\subsection{Model results}
\label{sec:results}

To study the viability of the four-fermion effective interaction
(\ref{eq:4fermi}) as a model for invariant mass and transverse
momentum dependence of the $t\bar t$ forward-backward
asymmetry observed by the CDF Collaboration~\cite{Aaltonen:2012it},
we performed scans over the coupling constants $g_{L,R},g'_{L,R}$,
for both $s$-channel colour octet and singlet
interactions, taking
into account the interference with the QCD contribution in the octet
case, both at the Born level and in one gluon emission.

The scans were restricted to coupling values between $+3$ and $-3$ TeV$^{-1}$,
to limit disagreement with the invariant mass distribution
of the cross section.  We also restricted the scans to two dimensions
by assuming equality of the light and heavy quark couplings,
$g'_L=g_L$ and $g'_R=g_R$.\footnote{In the octet case, we also
  performed a scan assuming $g'_L=-g_L$ and  $g'_R=-g_R$, which is
  different owing to the interference with QCD,  but the fits were
  less good.}     A step size of 0.4 TeV$^{-1}$, with the
reflection symmetry of the couplings, meant that 128 model points were
scanned.  At each model point, predictions were generated using $10^7$
weighted Monte Carlo phase space points, and $\chi^2$ values were
computed for fits to the CDF data.   We used the
MSTW2008 NLO (68cl) parton density function set~\cite{Martin:2009iq},
with the strong coupling evaluated at the top mass, taken to be
$m_t=173$ GeV.

\FIGURE{
  \centering\centerline{
  \includegraphics[scale=0.8]{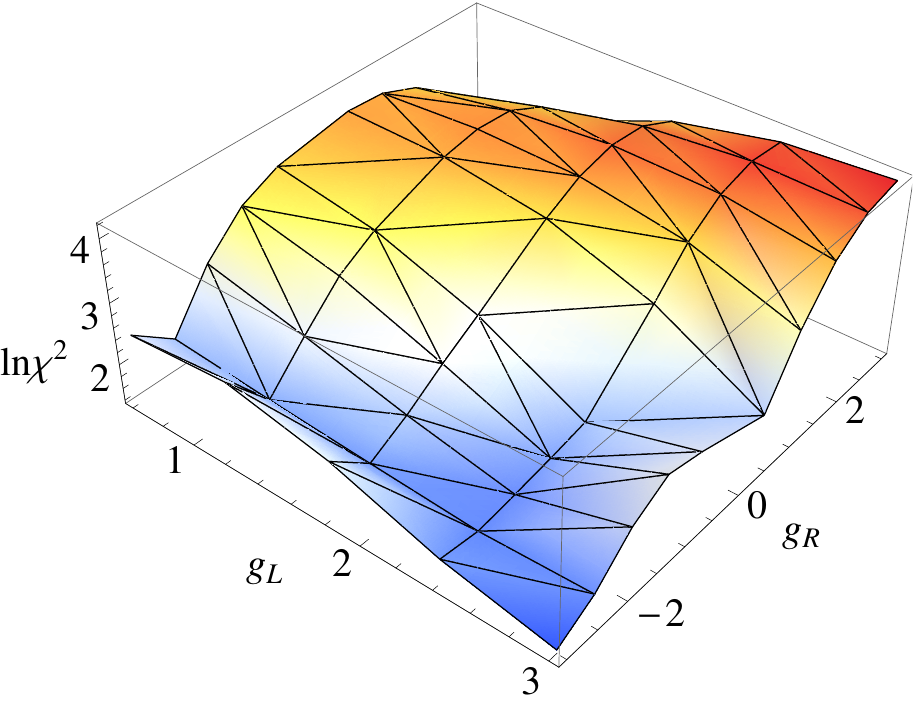}
  \includegraphics[scale=0.8]{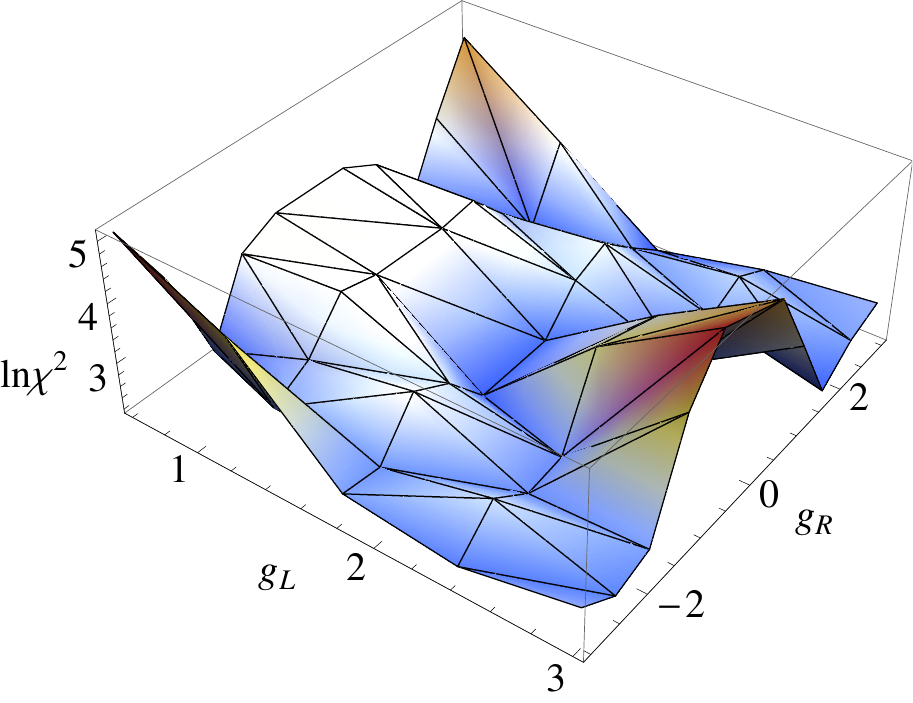}}
\vspace*{-5mm}
 \caption{\label{fig:scans}%
Values of $\ln\chi^2$ versus couplings $g_L=g'_L$ and
$g_R=g'_R$ (units of TeV$^{-1}$) 
for colour octet (left) and colour singlet (right)
four-fermion BSM interactions, compared to CDF data on the
transverse momentum dependence of the asymmetry.}}

Figure~\ref{fig:scans} shows for example the $\chi^2$ values for the fit to
the transverse momentum dependence of the asymmetry, assuming either
an $s$-channel colour octet (left) or singlet (right) four-fermion BSM
interaction.\footnote{More generally, one could allow for an arbitrary
  linear combination of octet and singlet contributions. Exchange of a
  colour triplet diquark in the $t$-channel, for example, leads to an
  equal admixture of octet and singlet in our effective theory. However, we shall see that
  the transverse momentum dependence qualitatively favours a pure octet.} Only the region $g_L>0$ is shown, on account of the
overall reflection symmetry of the couplings. We note, moreover, that
for the colour singlet interaction, there is approximate symmetry
under $g_R=g'_R\rightarrow -g_R=-g'_R$, at fixed $g_L=g'_L$.
Indeed, the diagrams we consider are
invariant under $g_R \rightarrow -g_R$, at fixed
$g_L$ and {\em vice versa}. This is easily understood: our diagrams
feature a single light quark line, along which helicity is conserved.
In the singlet case, there is no interference with QCD,  and so  the
sum of the powers of $g_L$ and $g_R$ in the matrix
element squared must be two. The only such term which would not
exhibit the claimed reflection symmetry in $g_L$ and $g_R$ separately
is $g_L g_R$. This term could only arise via interference between two
diagrams, one with a left-handed light quark line and one with a right-handed quark
line. But such diagrams cannot interfere since the external light quarks in
the two diagrams are in different helicity states.  The analogous
reflection symmetry in the heavy quark couplings $g'_R$ and $g'_L$
separately is broken by mass terms, which lead to helicity flips along
quark lines and which account for the
symmetry in Fig.~\ref{fig:scans}  (right) not being exact.  However,
as we show in the Appendix, the neglect of mass terms in the matrix
elements would not affect the qualitative features of our results.

Best-fit values of the couplings were found separately for the
invariant mass and transverse momentum dependences, and for a
simultaneous fit to both.  The results are summarized in Table~\ref{tab:fits}.
\begin{table}
  \begin{center}    
    \begin{tabular}{|c|c|rr|r|r|}
      \hline
      Model & Best fit & $g_L$ & $g_R$& $A_{\rm FB}(M_{tt})$
      &$A_{\rm FB}(p_{Ttt})$ \\
      \hline
 Octet &$A_{\rm FB}(M_{tt})$   & 2.2 & --0.2 & $\chi^2=1.7$ & $\chi^2=11.4$\\
    &$A_{\rm FB}(p_{Ttt})$  & 3.0 & --3.0 & 95.5 & 4.0\\
    & Both                      & 3.0 &     0.2 & 2.8 &   5.8\\
\hline
 Singlet &$A_{\rm FB}(M_{tt})$ & 2.2 & --0.6 & $\chi^2=1.5$ & $\chi^2=27.1$\\
    &$A_{\rm FB}(p_{Ttt})$ & 3.0 &     1.8 & 12.1 & 8.7\\
    & Both                     & 1.8 & --0.2 &  3.4 &  9.6\\
\hline
QCD LO & & & & $\chi^2=27.4$ & $\chi^2=45.6$\\
\hline
   \end{tabular}
  \end{center}
  \caption{
\label{tab:fits} Best fit couplings (in units of TeV$^{-1}$) and
$\chi^2$ values for different models and observables.
}
\end{table}

Fig.~\ref{fig:topasy_Afbm} shows the best-fit
model results for the asymmetry as a function of $t\bar t$ invariant
mass, for colour octet (left panel) and colour singlet (right
panel)  interactions, compared to the CDF data.  In both cases one
can obtain a good description of the data with reasonable values of
the BSM couplings.  However, the predictions for the dependence of the
asymmetry on the $t\bar t$  transverse momentum (Fig.~\ref{fig:topasy_Afbp}) are
then very different, with the colour octet giving much better,
albeit not perfect, agreement with the CDF data. In accord
with the expectations discussed above, the octet interaction leads to
an asymmetry that falls with increasing  transverse momentum, while
the singlet gives one that is larger and more constant, even slightly rising.

\FIGURE{
  \centering\centerline{
 \includegraphics[scale=0.35]{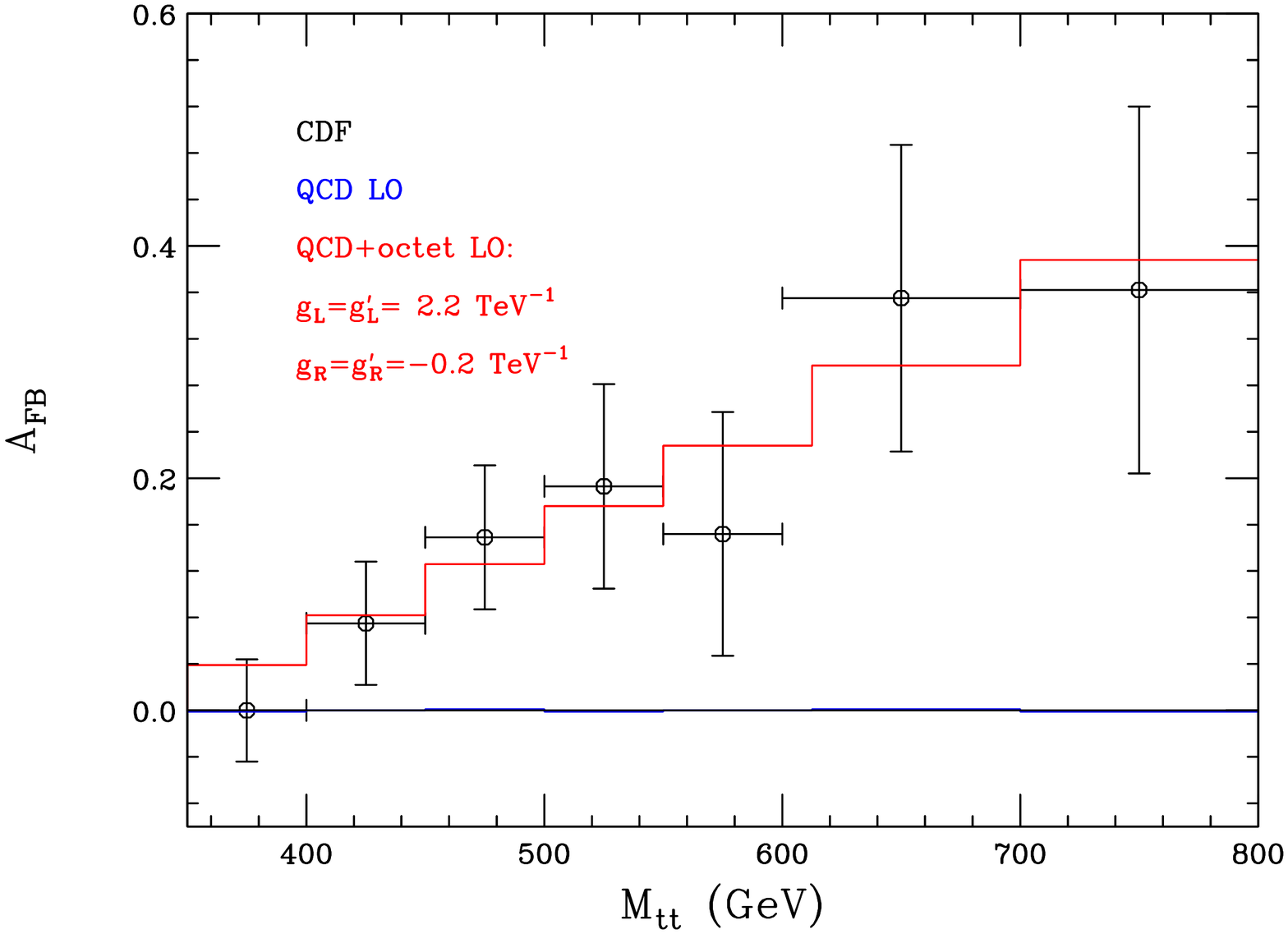}
\hspace*{-20mm}
  \includegraphics[scale=0.35]{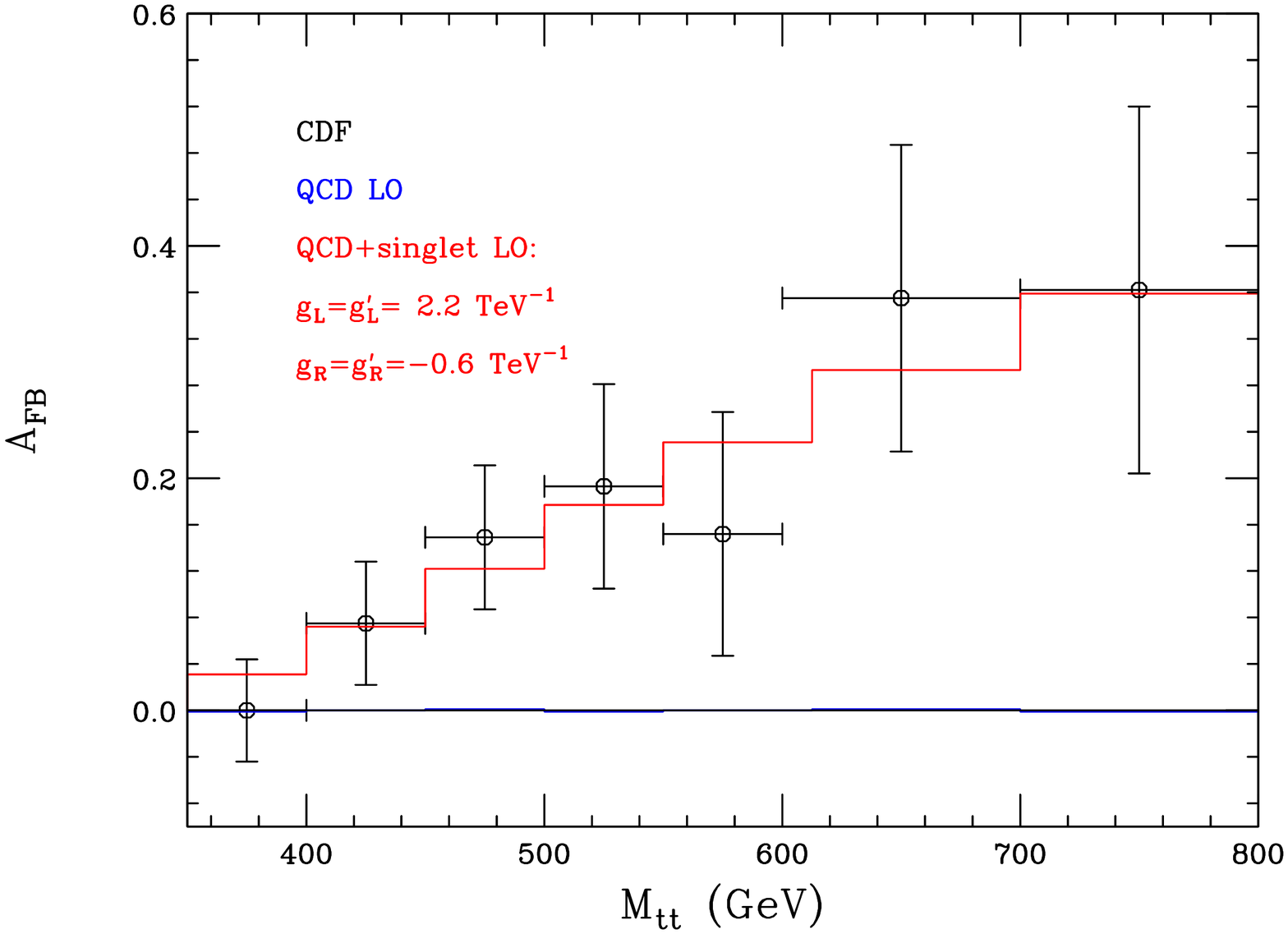}}
\vspace*{-10mm}
  \caption{\label{fig:topasy_Afbm}%
Best-fit model results for the forward-backward asymmetry as a function of the
$t\bar t$ invariant mass, for colour octet (left) and colour singlet
(right) four-fermion BSM interactions, compared to CDF data.}}

\FIGURE{
  \centering\centerline{
  \includegraphics[scale=0.35]{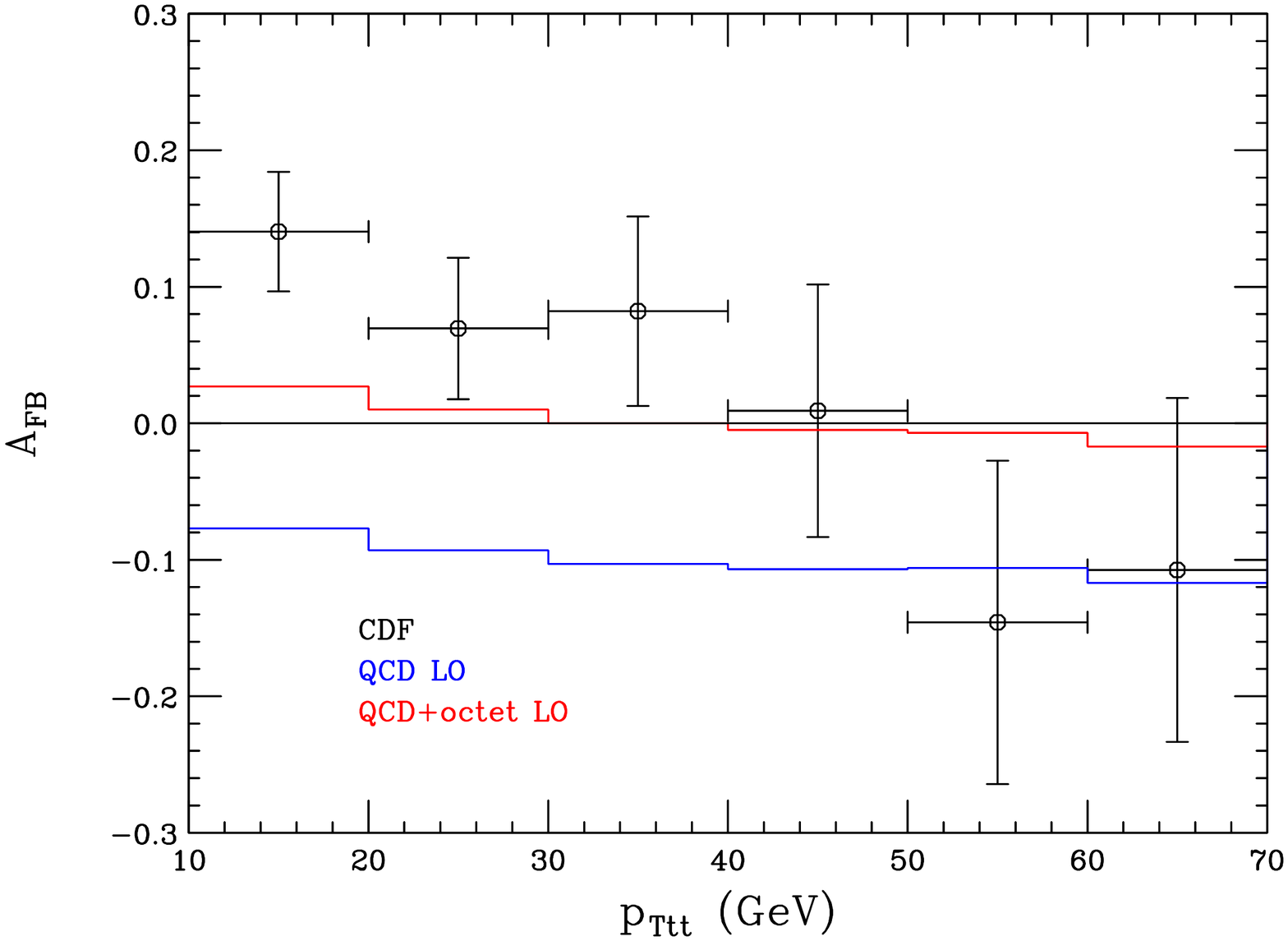}
\hspace*{-20mm}
  \includegraphics[scale=0.35]{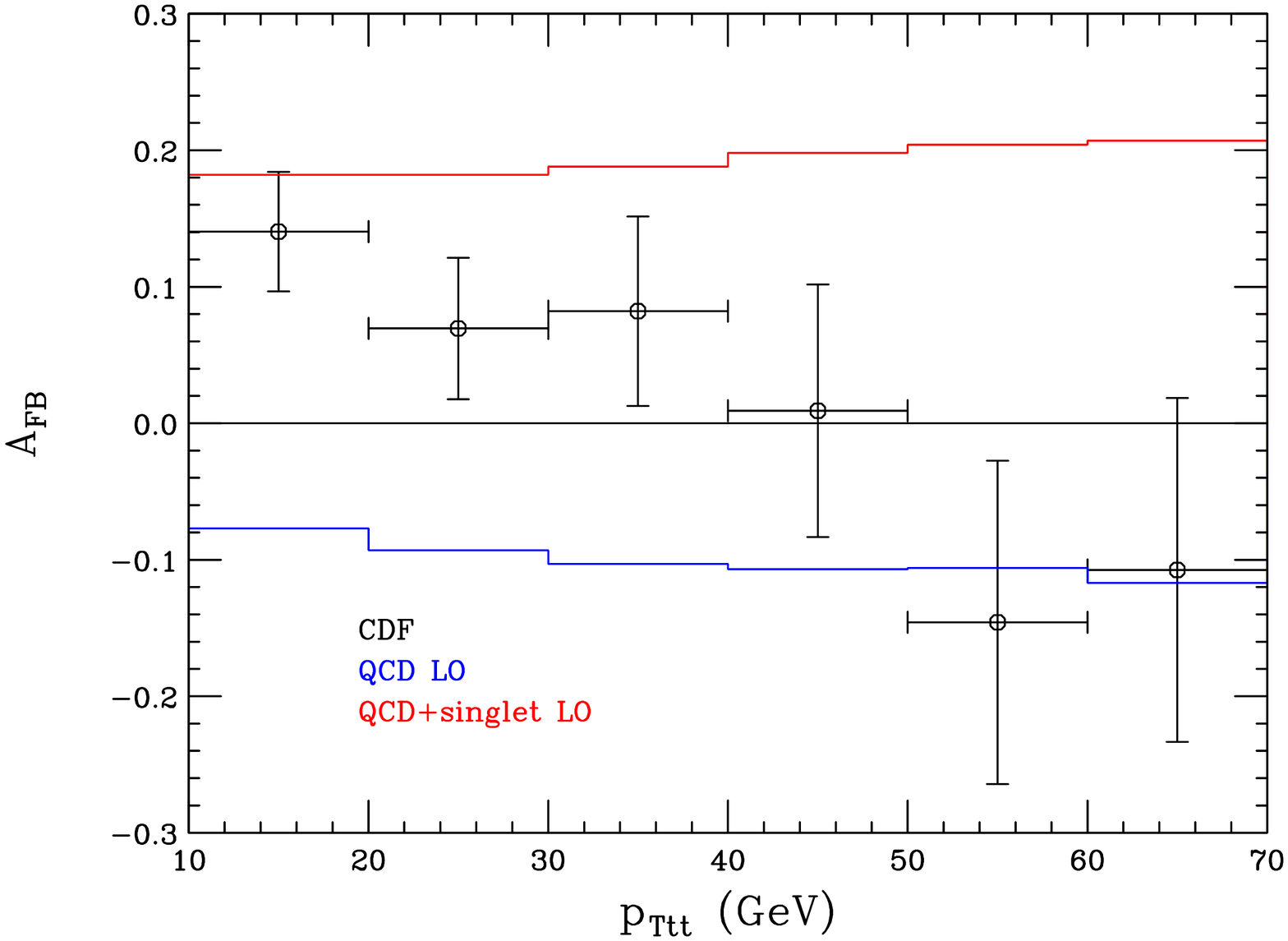}}
\vspace*{-10mm}
  \caption{\label{fig:topasy_Afbp}%
Model results for the forward-backward asymmetry as a function of the
$t\bar t$ transverse momentum, for colour octet (left) and colour singlet
(right) four-fermion BSM interactions, compared to CDF data.  Coupling
values as in Fig.~\ref{fig:topasy_Afbm}.}}

Although both the octet and singlet interactions can describe the
invariant mass dependence well, neither gives a very good description
of the separate forward and backward production cross sections,
Fig.~\ref{fig:topasy_sigfb}, as both involve a BSM contribution that
enhances the cross section at high mass.  Nevertheless the octet gives
significantly better agreement.

\FIGURE{
  \centering\centerline{
  \includegraphics[scale=0.35]{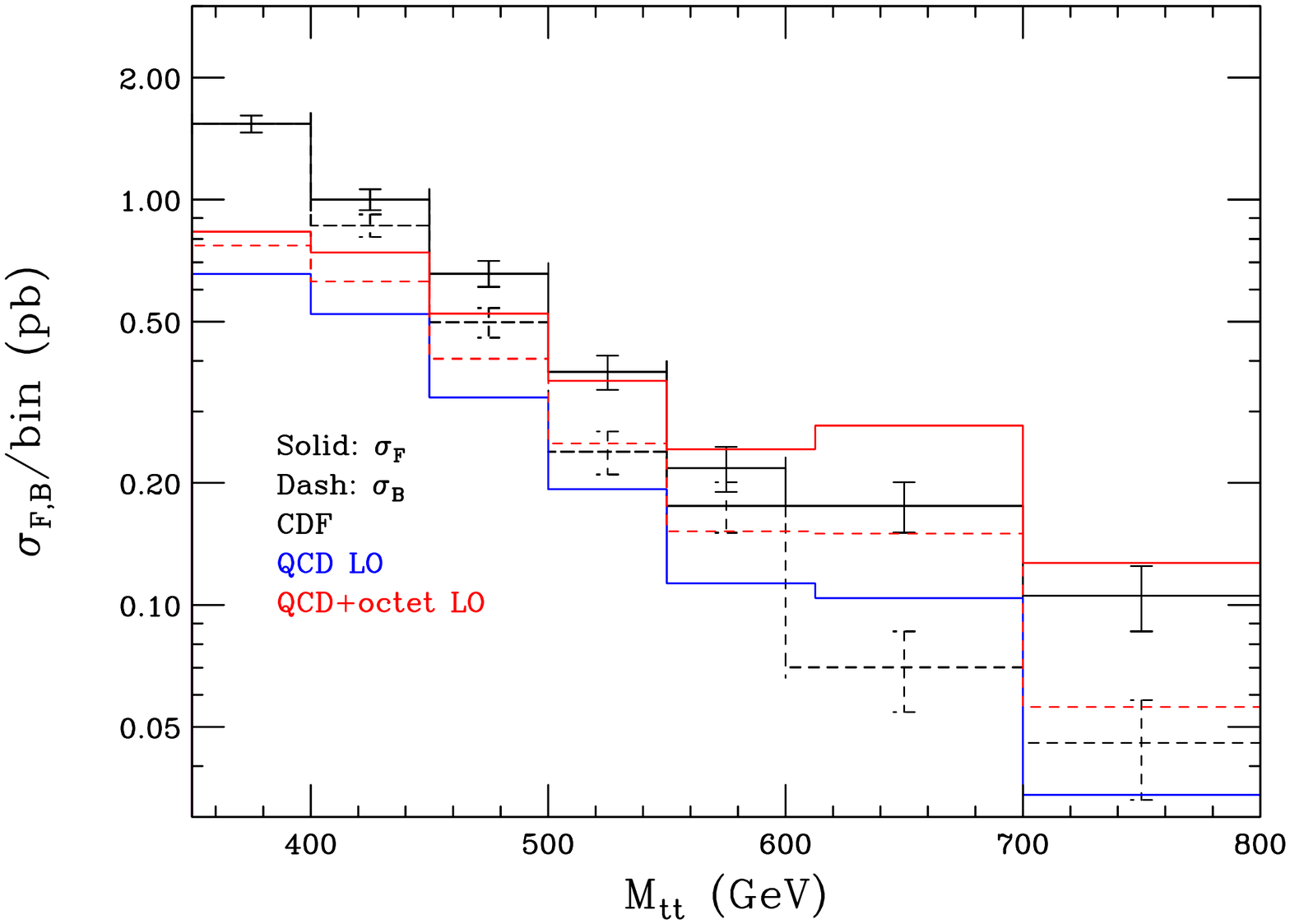}
\hspace*{-20mm}
  \includegraphics[scale=0.35]{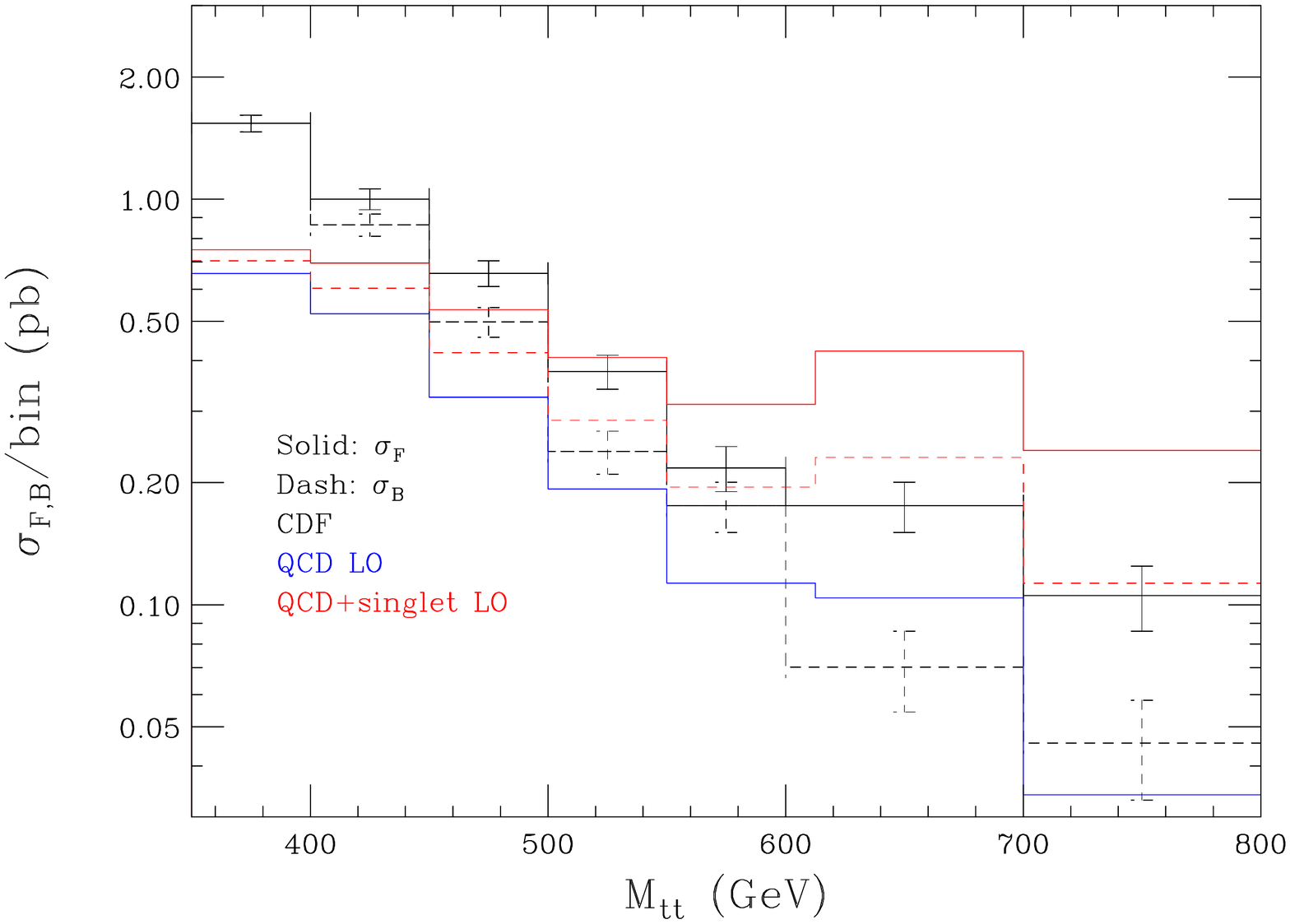}}
\vspace*{-10mm}
 \caption{\label{fig:topasy_sigfb}%
Model results for the forward (solid) and backward (dashed)  $t\bar t$
cross sections, for colour octet (left) and colour singlet (right)
four-fermion BSM interactions, compared to CDF data. Coupling
values as in Fig.~\ref{fig:topasy_Afbm}.}}

The fit of the singlet model to the transverse momentum dependence of
the asymmetry can be improved by making the couplings larger and more
left-right symmetric, but then the description of the invariant mass
dependence becomes much worse, as shown in Fig.~\ref{fig:topasy_Afbpm}.

\FIGURE{
  \centering\centerline{
  \includegraphics[scale=0.35]{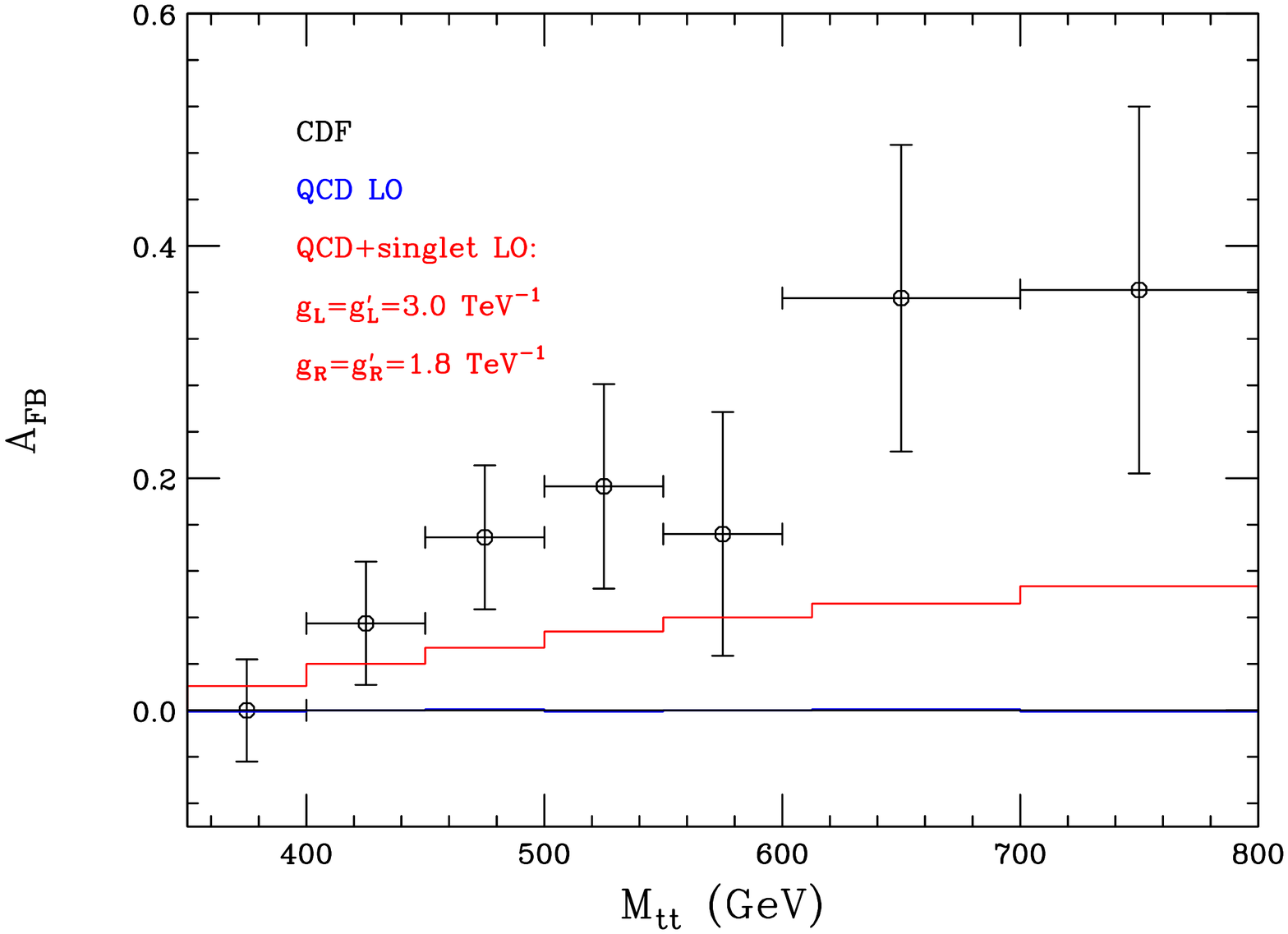}
\hspace*{-20mm}
  \includegraphics[scale=0.35]{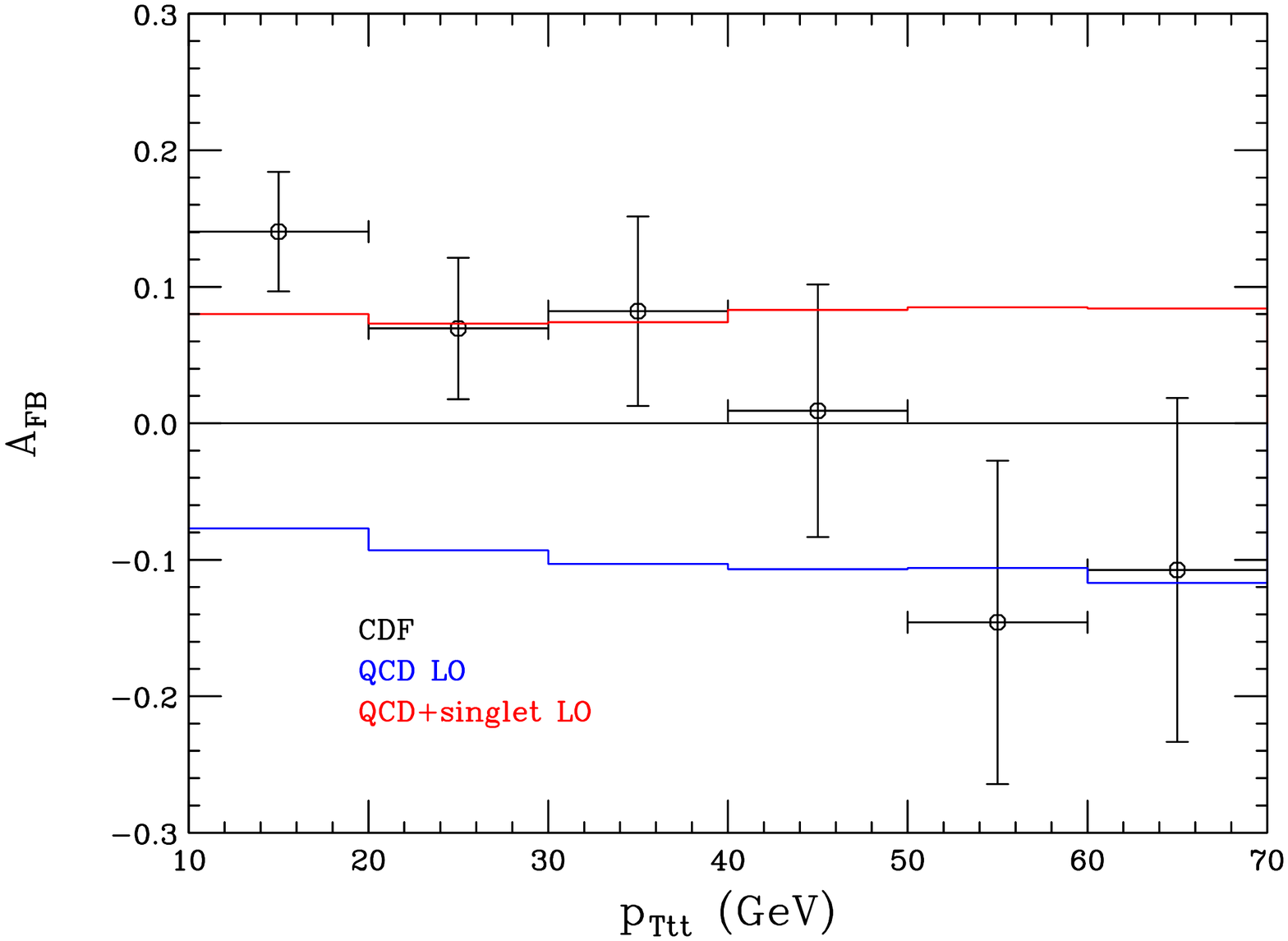}}
\vspace*{-10mm}
  \caption{\label{fig:topasy_Afbpm}%
Best-fit singlet model results for the forward-backward asymmetry as a function of the
$t\bar t$ transverse momentum, with the corresponding prediction for
the invariant mass dependence, compared to CDF data.}}

Overall, as other authors have noted~\cite{Zhang:2010dr,Degrande:2010kt,Blum:2011up,Delaunay:2011gv,AguilarSaavedra:2011vw,Hewett:2011wz}, the four-fermion BSM
interaction has problems fitting the asymmetry data, but does better
than leading-order QCD alone.  We have found that an $s$-channel colour octet
rather than singlet form of interaction is definitely preferred.  This
leaves open the possibility that QCD higher-order interactions could
explain the data equally well.

\section{Monte Carlo studies}
\label{sec:MC}
\subsection{Explicit new physics models}
We extend our discussion by investigating the inclusion of explicit
new resonances in conjunction with leading-order QCD. We focus on two
models corresponding to the colour-singlet and colour-octet effective interactions investigated up to this point. The first is a $Z'$ ($Z$-prime) colour-singlet vector boson whose
Lagrangian contains interactions of the form:
\beq 
(c_L^i \bar{q}_i \gamma^\mu P_L q_i + c_R^i\bar{q}_i \gamma^\mu P_R q_i)Z'_\mu\;,
\eeq
where $c^i_{L,R}$ are the couplings of the $Z'$ to the left- and
right-handed quarks of flavour $i$, and $Z'_\mu$ is the $Z'$
field. We will focus on the case where the only non-zero couplings are
those to the up and top quarks, and couplings of the same chirality
are equal ($c^i_L = c^j_L$, $c^i_R = c^j_R$, $i \neq j$). The second model we consider is that of a new
colour-octet vector boson, which we will refer to as `axigluon', $\tilde{G}$, whose
Lagrangian contains interactions of the form:
\beq
g_s [ \bar{q}_i T^A \gamma^\mu ( c_L^i P_L + c_R^i P_R) q_i + \bar{t} T^A
\gamma^\mu (c^t_L P_L + c_R^t P_R ) t ] \tilde{G}^A_\mu\;,  
\eeq
where $g_s$ is the strong coupling constant of QCD, $c^i_{L,R}$ are
the left- and right-handed couplings to $q_i\bar{q}_i$ (excluding the top
quark), $c^t_{L,R}$ are the left- and right-handed couplings to
$t\bar{t}$, the $T^A$ ($A\in \{1,8\}$) are the $SU(3)$ generators
in the adjoint representation, and $\tilde{G}^A_\mu$ is the
axigluon field. We will assume that the axigluon couplings to all the
quark flavours (including the top quark) of the same chirality are equal and denote them by $c_{L,R}$.

The resulting effective four-fermion interactions will contain the heavy boson
propagator. The propagator has a
$\sim 1/M^2$ dependence on the boson mass, allowing us to make the
identification $g \leftrightarrow c/M$ between the couplings of the effective theory and the
explicit models, in the large $M$ limit.  

\subsection{Monte Carlo implementation}
Both of the above models are available in the {\tt HERWIG++} event
generator~\cite{Bahr:2008pv, Arnold:2012fq}. {\tt HERWIG++} constructs all the possible leading-order
diagrams for $q\bar{q} \rightarrow t\bar{t}$, including the relevant
interference between the colour-octet $\tilde{G}$ and the QCD
diagrams. An angular-ordered shower is added on top of the resulting
matrix elements. Evidently, the finite gluon emission from internal
QCD gluons is not included in this calculation. We also note that in
Ref.~\cite{Skands:2012mm} it was found that {\tt HERWIG++}
underestimates the effects of QCD coherence in the asymmetry. 

A switch for limiting the shower to a single gluon emission either from the
initial- or final-state partons has been added to the event generator and will be publicly available in
the near future. In these studies we again use the
MSTW2008 NLO (68cl) parton density function set~\cite{Martin:2009iq}. We assume
perfect reconstruction of the top and anti-top quarks, ignoring experimental effects. 

\subsection{Monte Carlo results}
We performed fits to the CDF data equivalent to those described in
Section~\ref{sec:results}. The scans were restricted to coupling values $c_{L,R} \in [-50.0,
+50.0]$ in steps of width $2.0$. We considered heavy masses between
$M = 800$~GeV and $M = 2400$~GeV in steps of $400$~GeV. For each Monte Carlo point we generated $10^5$ events, either restricting the shower to one
gluon emission or with the full shower. Table~\ref{tab:fits_mc_oneg}
summarizes the results for the best fits for the one gluon emission
case, for different mass values. Table~\ref{tab:fits_mc_shower} summarizes the
equivalent results for the best fits for the full shower case.  We
also show the leading-order QCD result using the internal {\tt
  HERWIG++} matrix elements, for comparison. The
conclusions drawn from these results do not differ from those obtained
by the use of the effective theory: Good descriptions of the data for
 $A_{\mathrm{FB}}(M_{tt})$ and $A_{\mathrm{FB}}(p_{Ttt})$ can be
obtained in both the $Z'$ and $\tilde{G}$ models, for any
mass value. There is a tendency for the fits to
improve for larger boson masses and the full shower results fit the
CDF data better than those including only a single gluon emission. 

\begin{table}
  \begin{center}    
    \begin{tabular}{|c|c|rr|r|r|}
      \hline
      Model & Best fit & $c_L$ & $c_R$& $A_{\rm FB}(M_{tt})$
      &$A_{\rm FB}(p_{Ttt})$ \\
      \hline
 Axigluon &$A_{\rm FB}(M_{tt})$   & 16.0 & 8.0 & $\chi^2=3.8$ & $\chi^2=3.7$\\
    M = 800~GeV&$A_{\rm FB}(p_{Ttt})$  & 16.0 & 30.0 & 4.1 & 1.7\\
    & Both                      & 16.0 &   30.0 & 4.1 &   1.7\\
\hline
Axigluon &$A_{\rm FB}(M_{tt})$   & 16.0 & --6.0 & $\chi^2=1.4$ & $\chi^2=7.6$\\
    M = 1600~GeV&$A_{\rm FB}(p_{Ttt})$  & 14.0 & --30.0 & 2.9 & 1.4\\
    & Both                      & 14.0 &   --30.0 & 2.9 &   1.4\\
\hline
 Axigluon &$A_{\rm FB}(M_{tt})$   & 34.0 & --2.0 & $\chi^2=1.8$ & $\chi^2=3.8$\\
    M = 2400~GeV&$A_{\rm FB}(p_{Ttt})$  & 2.0 & --32.0 & 2.2 & 1.9\\
    & Both                      & 2.0 &   --32.0 & 2.2 &   1.9\\
\hline
\hline
 $Z'$ &$A_{\rm FB}(M_{tt})$ & 4.0 & 2.0 & $\chi^2=4.1$ & $\chi^2=18.0$\\
    M = 800~GeV &$A_{\rm FB}(p_{Ttt})$ & 32.0 &  --22.0 &  11.4 & 7.2\\
    & Both                     & 44.0 & 26.0 &   7.2 &  7.8\\
\hline
 $Z'$ &$A_{\rm FB}(M_{tt})$ & 32.0 & --12.0 & $\chi^2=1.5$ & $\chi^2=17.4$\\
    M = 1600~GeV &$A_{\rm FB}(p_{Ttt})$ & 10.0 &  22.0 & 5.3 & 5.8\\
    & Both                     & 4.0 & --10.0 &  3.0 &  6.8\\
\hline
 $Z'$ &$A_{\rm FB}(M_{tt})$ & 28.0 & --2.0 & $\chi^2=4.4$ & $\chi^2=10.5$\\
    M = 2400~GeV &$A_{\rm FB}(p_{Ttt})$ & 36.0 & --4.0 & 5.3 & 6.0\\
    & Both                     & 36.0 & --4.0 & 5.3  & 6.0  \\
\hline
\hline
 QCD 1-gluon &  &  &  & $\chi^2=21.6$ &
 $\chi^2=21.0$\\
\hline
   \end{tabular}
  \end{center}
  \caption{
\label{tab:fits_mc_oneg} Monte Carlo best-fit couplings for explicit new physics models of a $Z'$
and an axigluon and $\chi^2$ values for different masses and different
observables, restricting the shower to a single gluon emission.
}
\end{table}

\begin{table}
  \begin{center}    
    \begin{tabular}{|c|c|rr|r|r|}
            \hline
      Model & Best fit & $c_L$ & $c_R$& $A_{\rm FB}(M_{tt})$
      &$A_{\rm FB}(p_{Ttt})$ \\
      \hline
 Axigluon &$A_{\rm FB}(M_{tt})$   & 30.0 & 18.0 & $\chi^2=3.2$ & $\chi^2=11.8$\\
    M = 800~GeV&$A_{\rm FB}(p_{Ttt})$  & 28.0 & 50.0 &  4.6 & 11.2\\
    & Both                      & 12.0 &     22.0 & 3.7 &  11.3\\
\hline
Axigluon &$A_{\rm FB}(M_{tt})$   & 10.0 & --4.0 & $\chi^2=1.5$ & $\chi^2=7.9$\\
    M = 1600~GeV&$A_{\rm FB}(p_{Ttt})$  & 2.0 & 4.0 & 1.7 & 5.8 \\
    & Both                      & 2.0 & 4.0   & 1.7 & 5.8   \\
\hline
 Axigluon &$A_{\rm FB}(M_{tt})$   & 42.0 & 8.0 & $\chi^2=1.3$ & $\chi^2=6.5$\\
    M = 2400~GeV&$A_{\rm FB}(p_{Ttt})$  & 42.0 & 6.0 & 1.7  & 4.3  \\
    & Both                      & 42.0 &   8.0 & 1.3  & 6.5  \\
\hline
\hline
 $Z'$ &$A_{\rm FB}(M_{tt})$ & 30.0 & 16.0 & $\chi^2=4.6$ & $\chi^2=12.9$\\
    M = 800~GeV &$A_{\rm FB}(p_{Ttt})$ & 6.0 & --8.0 & 12.9 & 7.3\\
    & Both                     & 18.0 & 10.0 & 5.5 & 9.6  \\
\hline
 $Z'$ &$A_{\rm FB}(M_{tt})$ & 22.0 & --10.0 & $\chi^2=1.8$ & $\chi^2=9.7$\\
    M = 1600~GeV &$A_{\rm FB}(p_{Ttt})$ & 24.0 & --42.0   & 5.1 & 4.6\\
    & Both                     & 8.0 & --16.0 & 2.3  &  6.2 \\
\hline
 $Z'$ &$A_{\rm FB}(M_{tt})$ & 46.0 & --6.0 & $\chi^2=1.6$ & $\chi^2=7.8$\\
    M = 2400~GeV &$A_{\rm FB}(p_{Ttt})$ & 8.0 & 0.0   & 2.0 & 4.7\\
    & Both                     & 34.0 & --2.0  & 1.7  &  5.0 \\
\hline
\hline
 QCD shower & &  &  & $\chi^2=13.2$ &
 $\chi^2=13.8$\\
\hline
   \end{tabular}
  \end{center}
  \caption{
\label{tab:fits_mc_shower} Monte Carlo best-fit couplings for explicit new physics models of a $Z'$
and an axigluon and $\chi^2$ values for different masses and different
observables, with the full shower.}
\end{table}

In Fig.~\ref{fig:topasy_Afbm_MC} we show the best-fit model results for the asymmetry as a function of
$t\bar{t}$ invariant mass, for boson masses $M = 1600$~GeV and for the
full shower. The predictions for
the $t\bar{t}$ transverse momentum distribution (Fig.~\ref{fig:topasy_Afbp_MC}) exhibit
differences between the $Z'$ and $\tilde{G}$ similar to those
observed in Fig.~\ref{fig:topasy_Afbp} between the singlet-
and colour-octet effective four-fermion interactions: the axigluon leads to
an asymmetry that falls with increasing $p_{Ttt}$, whereas the singlet
gives one that is roughly constant. 

Additionally, we performed simultaneous fits to all four CDF
distribution, {\em i.\ e.} including
the forward and backward differential cross sections with respect to
the $t\bar{t}$ invariant mass. To obtain a fit that is better than leading-order QCD
alone, it is necessary for the heavy boson mass in both models to lie
in the region $M \gtrsim
1600$~GeV. The axigluon model gives a better fit to all
distributions, particularly to the forward and backward differential
cross sections, as was observed in Section~\ref{sec:results}. The
values of the couplings for the best overall fits to all distributions
are shown in Table~\ref{tab:fits_mc_overall} for both models, as well
as QCD at leading order. Due to the constraints coming from the
forward and backward cross sections, the couplings to the new bosons
are required to be small. 

\FIGURE{
  \centering\centerline{
 \includegraphics[scale=0.45]{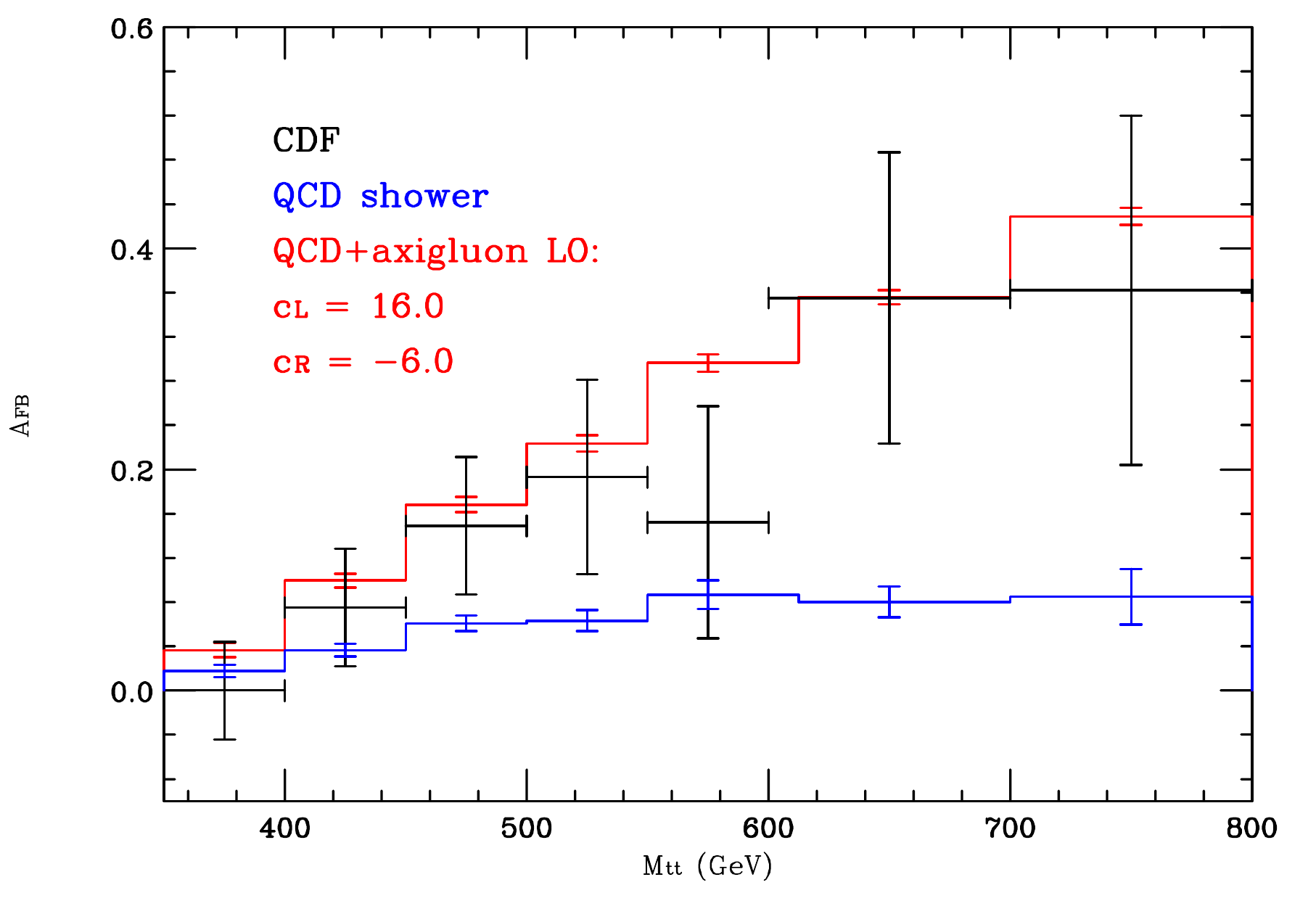}
\hspace*{-2mm}
  \includegraphics[scale=0.45]{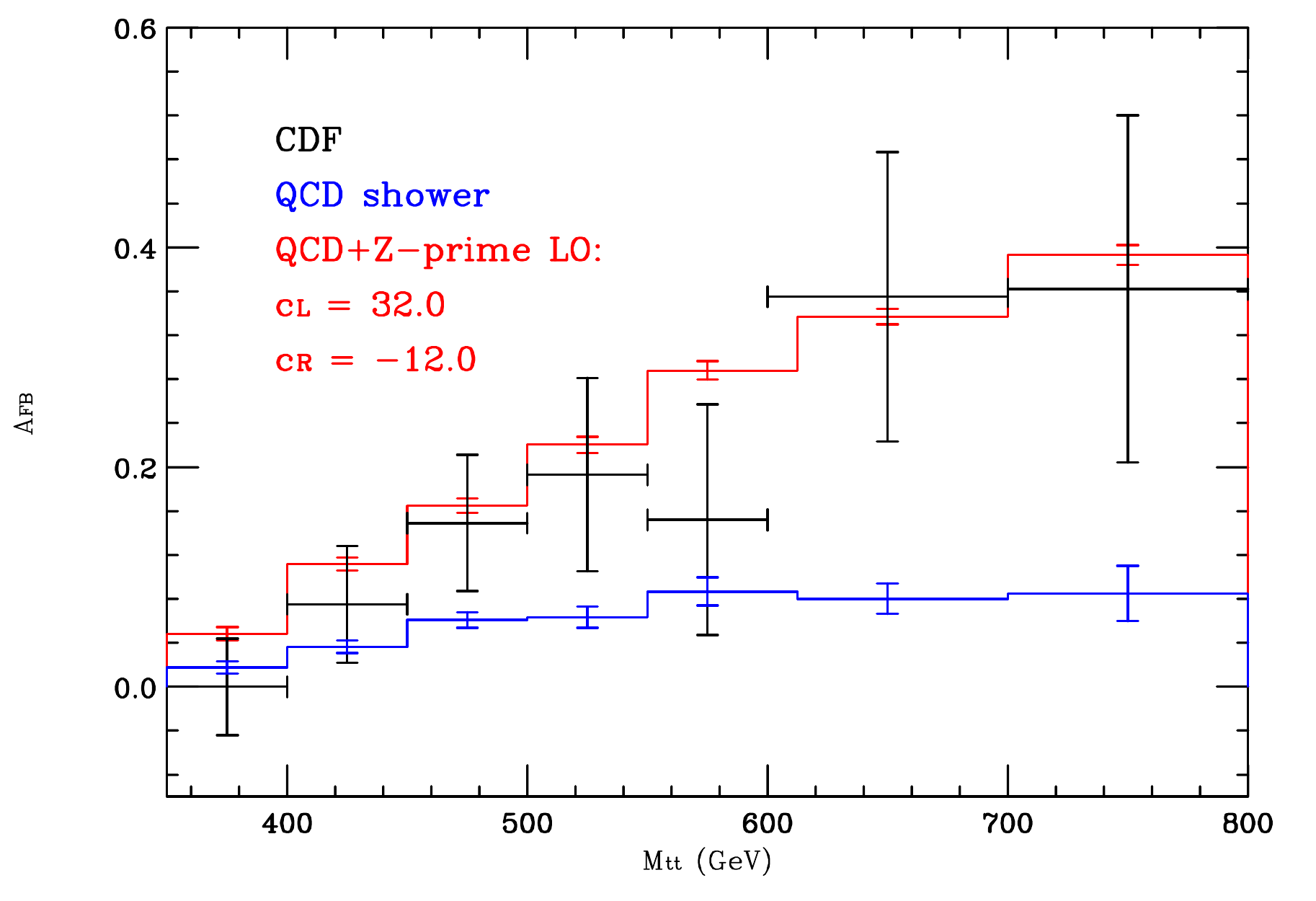}}
\vspace*{-2mm}
  \caption{\label{fig:topasy_Afbm_MC}
Monte Carlo best-fit model results for the forward-backward asymmetry as a function of the
$t\bar t$ invariant mass, for the colour-octet axigluon (left) and
colour-singlet $Z'$ (right), compared to CDF data. The QCD shower
result is included for comparison. The errors on the MC results are statistical.}}

\FIGURE{
  \centering\centerline{
  \includegraphics[scale=0.45]{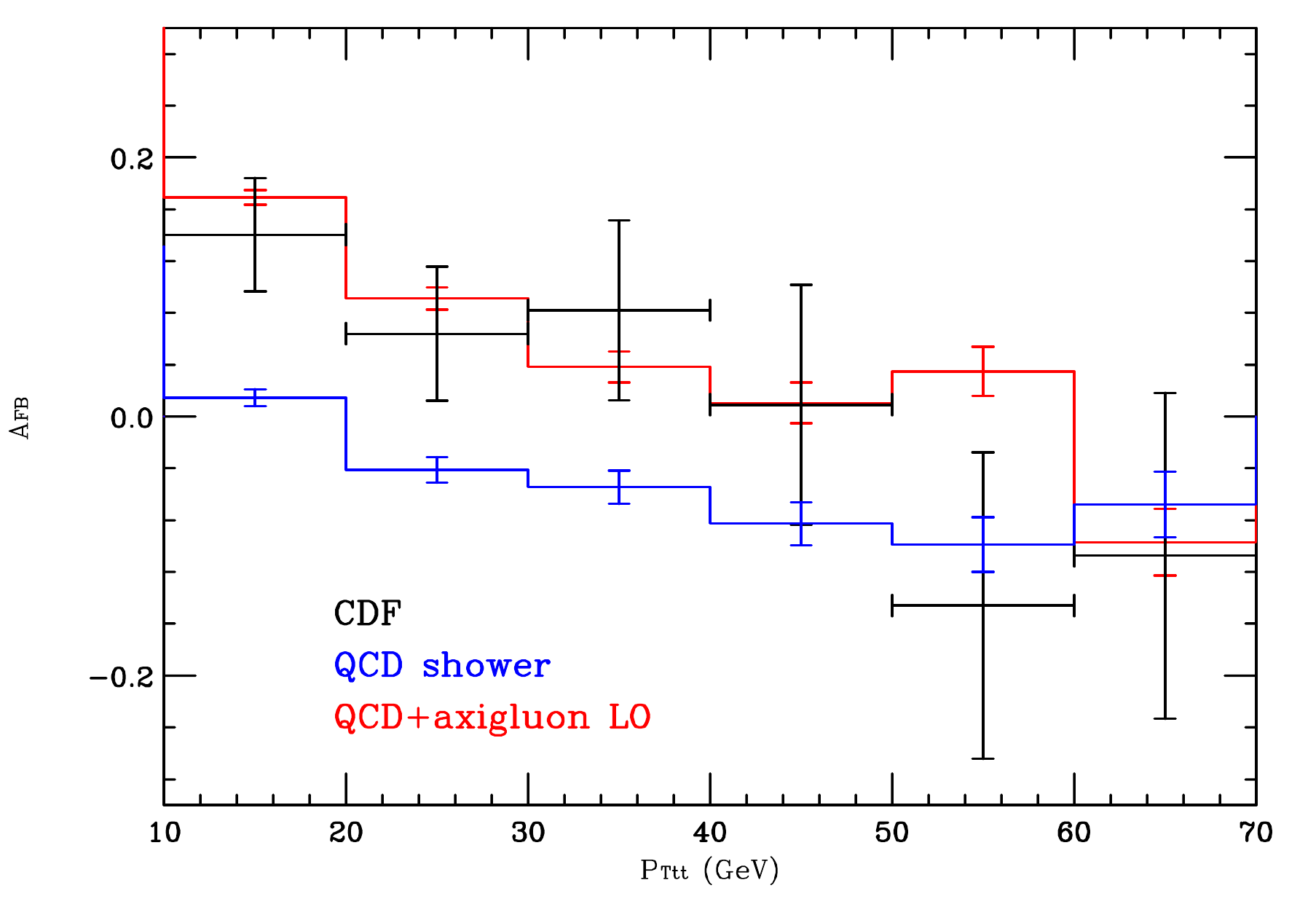}
\hspace*{-2mm}
  \includegraphics[scale=0.45]{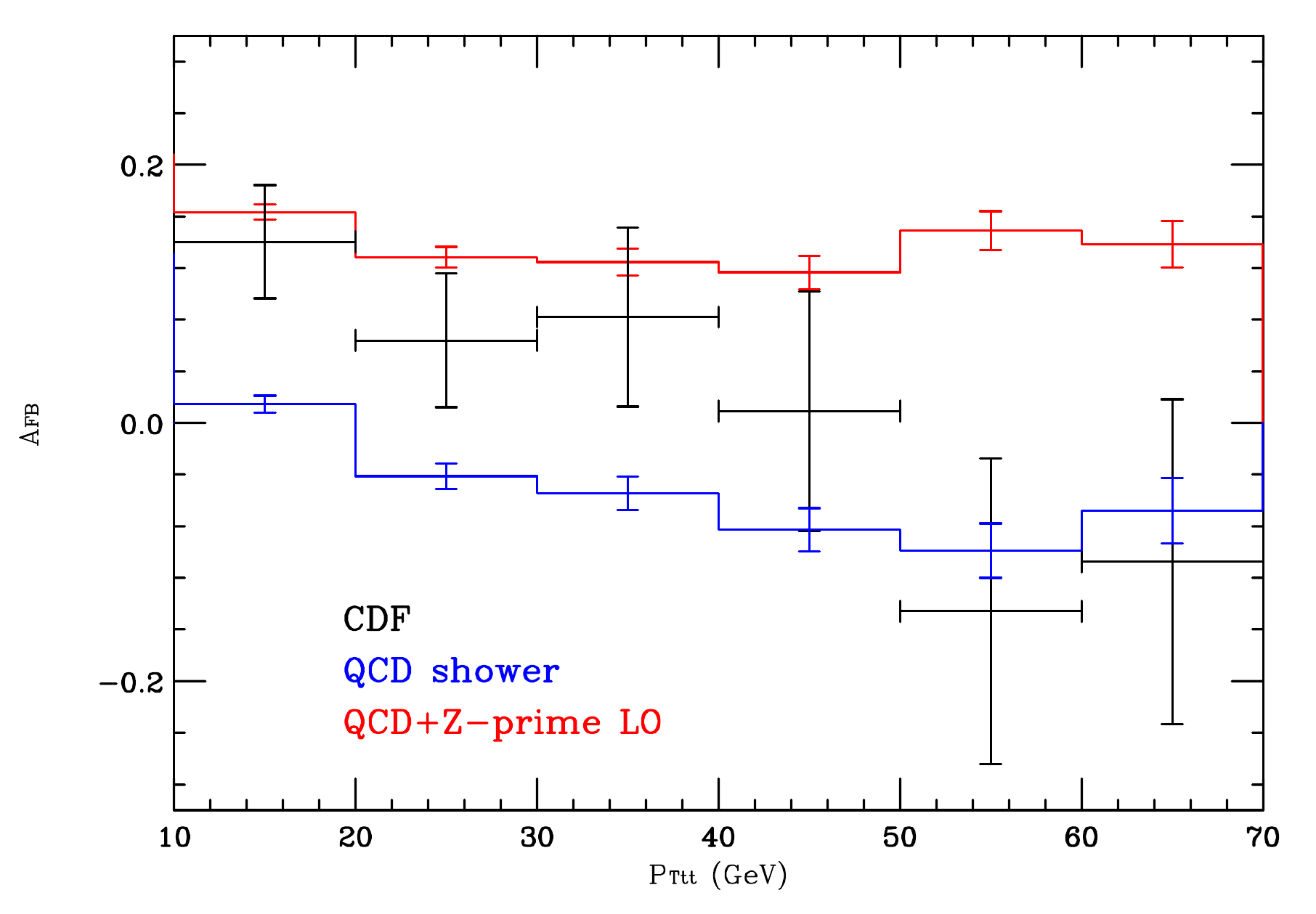}}
\vspace*{-2mm}
  \caption{\label{fig:topasy_Afbp_MC}
Monte Carlo best-fit model results for the forward-backward asymmetry as a function of the
$t\bar t$ transverse momentum, for the colour-octet axigluon (left) and
colour-singlet $Z'$ (right), compared to CDF data.  The QCD shower
result is included for comparison. The errors on the MC results are
statistical.}}

\begin{table}
  \begin{center}    
    \begin{tabular}{|c|c|rr|r|r|r|r|}
            \hline
Model & mass (GeV) & $c_L$ & $c_R$& $A_{\rm FB}(M_{tt})$ &$A_{\rm FB}(p_{Ttt})$ & $\sigma_{\rm F}(M_{tt})$ &$\sigma_{\rm B}(M_{tt})$  \\
\hline
Axigluon & 2000 & 4.0 & 2.0  & $\chi^2=5.3$ & $\chi^2 = 8.5  $ & $\chi^2 = 9.7$ & $\chi^2 = 27.8$ \\
\hline
$Z'$ & 2400 & 0.0 & 2.0  & $3.7$ & $8.9$ &
$20.8 $ & $62.3$  \\
\hline
 QCD shower & &  &  & $13.2$ &
 $13.8$ & 15.7 & 59.6\\
\hline
   \end{tabular}
  \end{center}
  \caption{
\label{tab:fits_mc_overall} Monte Carlo best-fit couplings for the explicit new physics models of a $Z'$
and an axigluon and $\chi^2$ values found during the full scan. Fitted
to all the differential distributions simultaneously, with the full shower. The QCD LO values are included
for comparison.}
\end{table}

\section{Conclusions}
\label{sec:conc}
Our objective in this paper was not to propose a particular model for
the top quark forward-backward asymmetry, nor indeed to advocate a BSM
origin of the asymmetry at all.  Our aim was rather to show, from
various viewpoints, that the dependence of the asymmetry on the
transverse momentum of the top pair provides additional information
not available from the inclusive asymmetry alone.  General features of
the QCD dynamics of gluon emission from the Born process $q\bar q\to
t\bar t$ ensure that mechanisms with the same inclusive asymmetry but
different colour structures will exhibit different transverse momentum
dependences.  In particular,  when the initial state is a colour
octet, the greater probability of gluon emission in backward top
production implies a negative correlation beween the asymmetry and the
transverse momentum of the top pair.  In the singlet state, on the
other hand, there is no correlation between the gluon emission
probability and the production angle and hence no such reduction of
the asymmetry.

We illustrated these effects first with an effective four-fermion
interaction that introduces an asymmetry at the Born level via
different left- and right-handed couplings to light and/or top
quarks. By computing the differential cross sections for one-gluon
emission in the octet and singlet states, including the interference
with pure QCD in the former, we showed that suitable choices of couplings
could give good fits to the inclusive asymmetry in both cases.
However, the corresponding transverse momentum dependences were then
quite different, in accordance with our qualitative expectations, the CDF
data favouring the octet state.

Whilst these conclusions seem qualitatively robust, it is important to
note that our calculation of the transverse momentum distribution is
effectively at the leading order of perturbation theory in QCD, and we
expect that there might be significant corrections at
next-to-leading-order, just as there are for pure
QCD~\cite{Dittmaier:2008uj,Melnikov:2011qx}.

We followed up this rather general fixed-order study with an
investigation of specific models in the approximate
all-orders framework provided by the {\tt HERWIG++} Monte Carlo event
generator.  Comparing results for a massive colour octet `axigluon'
and a singlet $Z'$ resonance decaying to $t\bar t$, both with variable
left- and right-handed couplings, interfaced to {\tt HERWIG++} parton
showers, we found qualitatively similar results to those from the
effective interaction.  The $Z'$ yields an asymmetry that is roughly
constant while that for the axigluon falls with increasing transverse
momentum, in better agreement with the CDF data.  However, the
axigluon mass needs to be high, above 1.6 TeV, and the couplings not
too strong, in order for the fit to the forward and backward cross
sections to be better than QCD alone.

\section*{Acknowledgements}
BG acknowledges
the support of the Science and Technology Facilities Council, the
Institute for Particle Physics Phenomenology, and King's College,
Cambridge, and thanks Nordita for hospitality and support during part
of this work. AP thanks Paolo Torrielli for useful
discussions and acknowledges support by the Swiss National Science
Foundation under contracts 200020-138206 and 200020-141360/1.
BW thanks Kirill Melnikov and Gavin Salam for helpful discussions,
acknowledges the support of a Leverhulme Trust Emeritus Fellowship,
and thanks the Pauli Institute at ETH/University of Zurich
for hospitality and support during part of this work. 

\appendix
\section{Amplitudes in the massless limit}

Neglecting masses, we use two-component spinor notation in the helicity basis
\beq
\gamma^\mu = \left(\begin{array}{cc} 0 & \sigma^\mu \\ \bsig^\mu
    & 0 \end{array}\right),
\eeq
where $\sigma^\mu = (1,\vec\sigma)$, $\bsig^\mu = (1,-\vec\sigma)$.
Left and right massless 2-spinors are represented as\footnote{The crucial sign factors are missing in
  some texts, which instead define, {\em e.g.} $k \rangle^\dagger
  \equiv [k$. The latter definition is inconsistent with the
  definition $ k \rangle [k \equiv 2\bsig_\mu  p^\mu_k$, as can
  be seen by comparing the traces $\mathrm{tr} \ k \rangle [k =
  \mathrm{tr} \ k \rangle k\rangle^\dagger = k\rangle^\dagger k
  \rangle $, which is positive-definite, and $\mathrm{tr} \ 2\bsig_\mu p^\mu_k
= 2p_k^0$, which can be of either sign in this all-states-outgoing
formalism. We also include a sign factor in our phase convention, choosing $u_R (p_k) = \mathrm{sign} (p_k^0) \;
i \sigma^2 u_L^* (p_k)$.}
\beeq
&&u_L(p_k) = v_R(p_k) = k]\;,\;\;\;
u_R(p_k) =  v_L(p_k) = k\rangle\;,\nonumber\\
&&u^\dag_L(p_k) = v^\dag_R(p_k) ={\rm sign}(p^0_k) \langle k\;,\;\;\;
u^\dag_R(p_k) =  v^\dag_L(p_k) = {\rm sign}(p^0_k) [k\;.
\eeeq
Then
\beq
k\equiv p_k^\mu\sigma_\mu= k]\langle k\,,\;\;\;
\bar k \equiv p_k^\mu\bsig_\mu= k\rangle[k\,.
\eeq
Spinor products satisfy
\beq
\langle jk\rangle = -\langle kj\rangle\,,\;\;\;
[jk]=-{\rm sign}(s_{jk})\langle jk\rangle^*=-[kj]
\eeq
where $s_{jk}=2p_j\cdot p_k$, so that
\beq
\spL{jk}\spR{kj} = {\rm sign}(s_{jk})|\spL{jk}|^2 = s_{jk}\,.
\eeq
Also
\beq
[j\sigma^\mu k\rangle =  {\rm sign}(s_{jk})\spM{k\bsig^\mu j}
= \spM{j\bsig^\mu k}^*\,.
\eeq
Then the relevant currents in the Born process (\ref{eq:4fermi}) are
\beeq
&&(\bar q\gamma^\mu q)_L = \spM{2\bsig^\mu 1}\,,\;\;\;
(\bar q\gamma^\mu q)_R = \spM{1\bsig^\mu 2}\,,\nn\\
&&(\bar q'\gamma^\mu q')_L = \spM{3\bsig^\mu 4}\,,\;\;\;
(\bar q'\gamma^\mu q')_R = \spM{4\bsig^\mu 3}\,.
\eeeq
Applying the Fierz identity
\beq
 \spM{1\bsig^\mu 2}\spM{3\bsig_\mu 4}=2\spL{31}\spR{24}\;,
\eeq
we have helicity amplitudes
\beeq
&&A_{LL}=2g_Lg'_L\spL{32}\spR{14}\,,\;\;\;
A_{RL}=2g_Rg'_L\spL{31}\spR{24}\,,\;\;\;\nn\\
&&A_{LR}=2g_Lg'_R\spL{42}\spR{13}\,,\;\;\;
A_{RR}=2g_Rg'_R\spL{41}\spR{23}\,.
\eeeq
Now $s_{31}=s_{24}=t$ and $s_{32}=s_{14}=u$. 
The BSM differential cross section is thus
\beq
\frac{d\sigma_{\rm BSM}}{dt} = \frac 1{16\pi s^2}
\left[(g_L^2\gp2_L+g_R^2\gp2_R)u^2+(g_L^2\gp2_R+g_R^2\gp2_L)t^2\right]\,,
\eeq
in agreement with the massless limit of eq.~(\ref{eq:dsdt}).
The QCD helicity amplitudes are the same, except that the couplings are
$g_{L,R}=g'_{L,R}=g_s$, there is a propagator factor of $1/s_{12}=1/s$,
and the cross section has a colour factor of $C_F/2N$.  It is
therefore equally straightforward to verify the other equations in
Sec.~\ref{sec:born} in the massless limit.
 
\subsection{One gluon emission}
We can represent the polarization of a gluon with momentum $p_5$ by
\beq
\beps_L = \frac{\bsig^\mu}{\sqrt 2}\frac{\spM{5\bsig_\mu
    r}}{\spR{5r}}\,,\;\;\;
\beps_R = \frac{\bsig^\mu}{\sqrt 2}\frac{\spM{r\bsig_\mu
    5}}{\spL{r5}}\,,
\eeq
where $p_r$ is a lightlike reference vector not along $p_5$.  Then for
emission from line 1 we have
\beeq
\spM{2\bsig^\mu 1}&\to&
\frac{g_s}{\spL{15}\spR{51}}\spM{2\bsig^\mu(1+5)\beps 1}\nn\\
&=&\frac{g_s}{\spL{15}\spR{51}}(\spM{2\bsig^\mu 1}\spM{1\beps 1}
+\spM{2\bsig^\mu 5}\spM{5\beps 1})\,.
\eeeq
Now, using Fierz again,
\beeq
\spM{1\beps_L 1} = \sqrt 2\spL{51}\spR{1r}/\spR{5r}\,,\;\;\;\;
&&\spM{1\beps_R 1} = \sqrt 2\spL{r1}\spR{15}/\spL{r5}\,,\nn\\
\spM{5\beps_L 1} = \sqrt 2\spL{55}\spR{1r}/\spR{5r}=0\,,\;
&&\spM{5\beps_R 1} = \sqrt 2\spL{r5}\spR{15}/\spL{r5}
=\sqrt 2\spR{15}\,.
\eeeq
Hence, denoting the amplitude for emission of a gluon of helicity
$h''$ from line $j$ by
$A^{(j)}_{hh',h''}$,
\beeq\label{eq:Ldef}
A^{(1)}_{LL,L} &=&\sqrt 2
g_Lg'_Lg_s\frac{\spR{1r}}{\spR{15}\spR{5r}}\spM{2\bsig^\mu 1}\spM{3\bsig_\mu
  4}\nn\\
&=&2\sqrt 2
g_Lg'_Lg_s\frac{\spL{32}\spR{14}\spR{1r}}{\spR{15}\spR{5r}}\equiv
2\sqrt 2 g_Lg'_L g_s L(1234)\,.
\eeeq
Similarly
\beeq\label{eq:Rdef}
A^{(1)}_{LL,R} &=&\sqrt 2
g_Lg'_L\frac{g_s}{\spL{51}}\left(
\frac{\spL{r1}}{\spL{r5}}\spM{2\bsig^\mu1}+
\spM{2\bsig^\mu5}\right)\spM{3\bsig_\mu 4}\nn\\
&=&2\sqrt 2
g_Lg'_Lg_s\frac{\spL{32}}{\spL{51}}\left(\frac{\spL{r1}}{\spL{r5}}\spR{14}
+\spR{54}\right)\equiv 2\sqrt 2 g_Lg'_L g_s R(1234)\,.
\eeeq
For emission from line 2
\beeq
\spM{2\bsig^\mu 1}&\to&
\frac{g_s}{\spL{25}\spR{52}}\spM{2\beps(2+5)\bsig^\mu 1}\nn\\
&=&\frac{g_s}{\spL{25}\spR{52}}(\spM{2\bsig^\mu 1}\spM{2\beps 2}
+\spM{5\bsig^\mu 1}\spM{2\beps 5})
\eeeq
and
\beeq
\spM{2\beps_L 2} = \sqrt 2\spL{52}\spR{2r}/\spR{5r}\,,\;\;\;\;
&&\spM{2\beps_R 2} = \sqrt 2\spL{r2}\spR{25}/\spL{r5}\,,\nn\\
\spM{2\beps_L 5} = \sqrt 2\spL{52}\,,\;
&&\spM{2\beps_R 5} =0\,.
\eeeq
Therefore
\beeq
A^{(2)}_{LL,L} &=&2\sqrt 2
g_Lg'_Lg_s\frac{\spR{14}}{\spR{25}}\left(
\frac{\spR{2r}}{\spR{5r}}\spL{32}+
\spL{35}\right)=2\sqrt 2 g_Lg'_Lg_s R^*(2143)\,,\nn\\
A^{(2)}_{LL,R} &=&2\sqrt 2
g_Lg'_Lg_s\frac{\spR{14}\spL{32}\spL{r2}}{\spL{52}\spL{r5}}=2\sqrt 2 g_Lg'_Lg_s
L^*(2143)\,.
\eeeq
In fact, all the contributions to the helicity amplitudes can be
expressed in terms of the two functions $L$ and $R$, defined by
Eqs.~(\ref{eq:Ldef}) and (\ref{eq:Rdef}) respectively, as shown in
Table~\ref{tab:helicity}.

\begin{table}
  \begin{center}    
    \begin{tabular}{|c|c|c|c|c|c|c|}
      \hline
      $hh',h''$& 1 & 2 & 3 & 4 & $g$ & Term \\
      \hline
 $LL,L$ & $L(1234)$ & $R^*(2143)$ & $R^*(3412)$ & $L(4321)$ &
 $G(1234)$ & $s_{14}^2$\\
 $LL,R$ & $R(1234)$ & $L^*(2143)$ & $L^*(3412)$ & $R(4321)$ &
 $G^*(2143)$ & $s_{23}^2$\\
 $LR,L$ & $L(1243)$ & $R^*(2134)$ & $L(3421)$ & $R^*(4312)$ &
 $G(1243)$ & $s_{24}^2$\\
 $LR,R$ & $R(1243)$ & $L^*(2134)$ & $R(3421)$ & $L^*(4312)$ &
 $G^*(2134)$ & $s_{13}^2$\\
 $RL,L$ & $R^*(1243)$ & $L(2134)$ & $R^*(3421)$ & $L(4312)$ &
 $G(2134)$ & $s_{13}^2$\\
 $RL,R$ & $L^*(1243)$ & $R(2134)$ & $L^*(3421)$ & $R(4312)$ &
 $G^*(1243)$ & $s_{24}^2$\\
 $RR,L$ & $R^*(1234)$ & $L(2143)$ & $L(3412)$ & $R^*(4321)$ &
 $G(2143)$ & $s_{23}^2$\\
 $RR,R$ & $L^*(1234)$ & $R(2143)$ & $R(3412)$ & $L^*(4321)$ &
 $G^*(1234)$ & $s_{14}^2$\\
 \hline
    \end{tabular}
  \end{center}
  \caption{
\label{tab:helicity}Helicity amplitudes for one gluon emission.
Column $i$ shows the function for emission from line $i$,
to be multiplied by the coupling factor $g_h g'_{h'}g_s$. Column $g$ refers
to the QCD emission from the internal gluon line.  The final column
shows the associated term in Eq.~(\protect\ref{eq:singlet1g}).
}
\end{table}

The colour factor for the various contributions to the matrix element
squared are given in Table~\ref{tab:colour}.
 For the colour-singlet case the matrix elements squared thus
take the form
\beq
{\cal M}^{\rm sing}_{hh',h''} = C_F\left(|A^{(1)}_{hh',h''}-A^{(2)}_{hh',h''}|^2
+|A^{(3)}_{hh',h''}-A^{(4)}_{hh',h''}|^2\right)\,.
\eeq
Consider for example the case $hh',h''=LL,L$.  We have
\beeq
&&A^{(1)}_{LL,L}-A^{(2)}_{LL,L} = 2\sqrt 2 g_Lg'_Lg_s\left[L(1234)-R^*(2143)\right] \nn\\
&&=2\sqrt 2
g_Lg'_Lg_s\frac{\spR{14}}{\spR{15}\spR{25}\spR{5r}}
\left[\left(\spR{25}\spR{1r}
-\spR{15}\spR{2r}\right)\spL{32}
-\spR{15}\spR{5r}\spL{35}\right]\,.
\eeeq
Applying the Schouten identity
\beq
\spR{ij}\spR{k\ell}+\spR{ik}\spR{\ell j}+\spR{i\ell}\spR{jk}=
\spL{ij}\spL{k\ell}+\spL{ik}\spL{\ell j}+\spL{i\ell}\spL{jk}=0
\eeq
we have
\beq
\spR{25}\spR{1r}-\spR{15}\spR{2r}=\spR{21}\spR{5r}
\eeq
so the dependence on the reference vector $r$ cancels and we find
\beq
A^{(1)}_{LL,L}-A^{(2)}_{LL,L} = 
2\sqrt 2
g_Lg'_Lg_s\frac{\spL{34}\spR{14}^2}{\spR{15}\spR{25}}\,.
\eeq
Here we have used momentum conservation to write
$\spL{32}\spR{21}+\spL{35}\spR{51}=\spL{34}\spR{14}$.
Similarly
\beeq
A^{(3)}_{LL,L}-A^{(4)}_{LL,L} &=& 2\sqrt 2
g_Lg'_Lg_s\left[R^*(3412)-L(4321)\right]\nn\\
&=&2\sqrt 2
g_Lg'_Lg_s\frac{\spL{12}\spR{14}^2}{\spR{45}\spR{35}}\,.
\eeeq
Taking the square moduli gives
\beeq \label{eq:bsm1}
|A^{(1)}_{LL,L}-A^{(2)}_{LL,L}|^2 &=& 8 g^2_L\gp2_Lg^2_s
s_{14}^2\frac{s_{34}}{s_{15}s_{25}}\,,\nn\\
|A^{(3)}_{LL,L}-A^{(4)}_{LL,L}|^2 &=& 8 g^2_L\gp2_Lg^2_s
s_{14}^2\frac{s_{12}}{s_{35}s_{45}}\;.
\eeeq
The other helicity amplitudes give similar contributions, with the
factor of $s_{14}^2$ replaced by the corresponding term in the final
column of Table~\ref{tab:helicity}.  Thus the overall spin-averaged
matrix element squared is
\beeq\label{eq:singlet1g}
\overline{\sum}{\cal M}^{\rm sing} = 2C_F g_s^2&&\Bigl[
(g^2_L\gp2_L+g^2_R\gp2_R)(s_{14}^2+s_{23}^2)\nn\\
&&+
(g^2_L\gp2_R+g^2_R\gp2_L)(s_{13}^2+s_{24}^2)\Bigr]
\left(\frac{s_{34}}{s_{15}s_{25}}+\frac{s_{12}}{s_{35}s_{45}}\right)\,.
\eeeq

In the colour octet (adjoint representation) case, the reference
vector dependence cancels for any value of $N$.  Using the colour
factors in Table~\ref{tab:colour},  in place of (\ref{eq:singlet1g}) we have
\beeq\label{eq:octet1g}
\overline{\sum}{\cal M}^{\rm oct} = 2g_s^2&&\Bigl[
(g^2_L\gp2_L+g^2_R\gp2_R)(s_{14}^2+s_{23}^2)\nn\\
&&+
(g^2_L\gp2_R+g^2_R\gp2_L)(s_{13}^2+s_{24}^2)\Bigr]
\sum_{i<j} C_{ij} \frac{s_{k\ell}}{s_{i5}s_{j5}}\,,
\eeeq
where $k,\ell\ne i,j,5$.

\subsection{QCD amplitudes}
The QCD helicity amplitudes for one gluon emission are as in the octet
case above, except for the following changes:
\begin{itemize}
\item The couplings are $g_{L,R}=g'_{L,R}=g_s$.
\item The gluon propagator gives a factor of $1/s_{34}$ in
  $A^{(1,2)}_{hh',h''}$ and $1/s_{12}$ in $A^{(3,4)}_{hh',h''}$.
\item There is an extra contribution $A^{(g)}_{hh',h''}/s_{12}s_{34}$
  from the diagram with a gluon emitted from the gluon propagator.
\end{itemize}
For $hh'=LL$, the numerator of the extra contribution is
\beeq
A^{(g)}_{LL,L}&=&2\sqrt 2 g^3_s\spR{14}\left(
\frac{\spL{53}\spR{3r}+\spL{54}\spR{4r}}{\spR{5r}}\spL{32}-\spL{35}\spL{52}\right)
\equiv 2\sqrt 2 g_s^3 G(1234)\,,\nn\\
A^{(g)}_{LL,R}&=&2\sqrt 2 g^3_s\spL{32}\left(
\frac{\spL{r3}\spR{35}+\spL{r4}\spR{45}}{\spL{r5}}\spR{14}-\spR{15}\spR{54}\right)
= 2\sqrt 2 g_s^3 G^*(2143)\;,
\eeeq
with permutations for the other helicities as shown in
Table~\ref{tab:helicity}.

The colour factors are such that emissions from lines 3 and 4 receive
an extra contribution $A^{(g)}_{hh',h''}/2s_{12}s_{34}$, while those
from 1 and 2 receive the opposite contribution.  Therefore the
extra contribution cancels in the $i,j=1,2$ and $3,4$ terms of
Eq.~(\ref{eq:octet1g}) but not in the others.  Consider for example
the term involving
\beq \label{eq:qcd1}
A^{(1)}_{LL,L}/s_{34}-A^{(4)}_{LL,L}/s_{12} - A^{(g)}_{LL,L}/s_{12}s_{34}
=2\sqrt 2\frac{g_s^3}{s_{12}s_{34}}\left[L(1234)s_{12}-L(4321)s_{34}-G(1234)\right]\,.
\eeq
Now $s_{12}=s_{34}+s_{35}+s_{45}$, so this is equal to
\beq
2\sqrt 2\frac{g_s^3}{s_{12}s_{34}}\left\{L[(1234)-L(4321)]s_{34}
+L(1234)(s_{35}+s_{45})-G(1234)\right\}\,.
\eeq
We want to show that this is independent of the reference vector
$p_r$.  The first term in the curly bracket is independent, by the
argument in the previous subsection.  The remainder involves
\beeq
L(1234)(s_{35}+s_{45})-G(1234)
=\frac{\spL{32}\spR{14}}{\spR{15}\spR{5r}}
\biggl\{\spR{1r}(\spL{53}\spR{35}+\spL{54}\spR{45})
&-&\spR{15}(\spL{53}\spR{3r}+\spL{54}\spR{4r})\biggr\}\nn\\
&+&\spL{35}\spL{52}\spR{14}\,.
\eeeq
Applying the Schouten identity, the $r$-dependence cancels and we find
\beq
L(1234)(s_{35}+s_{45})-G(1234)
=\frac{\spR{14}}{\spR{15}}(\spL{32}\spL{25}\spR{21}+\spL{35}\spL{52}\spR{15})\,,
\eeq
where we have used momentum conservation to write 
$\spL{53}\spR{31}+\spL{54}\spR{41}=\spL{25}\spR{21}$.
Similarly, the extra term cancels the $r$-dependence of the other
terms in which it appears.  In fact, it ensures that the simple
correspondence between the helicity contributions and the terms in the
last column of Table~\ref{tab:helicity} remains valid.  Corresponding
to Eq.~(\ref{eq:octet1g}) we have
\beq\label{eq:QCD1g}
\overline{\sum}{\cal M}^{\rm QCD} = 4\frac{g_s^6}{s_{12}s_{34}}
\left(s_{14}^2+s_{23}^2+s_{13}^2+s_{24}^2
\right)\sum_{i<j} C_{ij} \frac{s_{ij}}{s_{i5}s_{j5}}\,.
\eeq

\subsection{Interference}
Given the helicity amplitudes listed above, it is straightforward
to combine the QCD and octet four-fermion contributions taking into
account their interference.  The combined amplitudes take the form
(showing the couplings explicitly)
\beeq
&&A^{(i)}_{hh',h''}\left(g_h g'_{h'}+\frac{g_s^2}{s_{34}}\right)
-A^{(g)}_{hh',h''}\frac{g_s^2}{2s_{12}s_{34}}\;\;(i=1,2)\nn\\
&&A^{(i)}_{hh',h''}\left(g_h g'_{h'}+\frac{g_s^2}{s_{12}}\right)
+A^{(g)}_{hh',h''}\frac{g_s^2}{2s_{12}s_{34}}\;\;(i=3,4)\,.
\eeeq
Consider first the $i=1,2$ contribution:
\beeq \label{eq:int1}
&&\overline{\sum}\left|g_hg'_{h'}(A^{(1)}_{hh',h''}-A^{(2)}_{hh',h''})
+\frac{g_s^2}{s_{12}s_{34}}
(A^{(1)}_{hh',h''}s_{12}-A^{(2)}_{hh',h''}s_{34})\right|^2\\
&=&\overline{\sum}\left\{|A^{\rm oct}|^2 +|A^{\rm QCD}|^2
+2\frac{g_hg'_{h'}g_s^2}{s_{12}s_{34}}\,
{\rm Re}\left[(A^{(1)}_{hh',h''}-A^{(2)}_{hh',h''})^*
(A^{(1)}_{hh',h''}s_{12}-A^{(2)}_{hh',h''}s_{34})\right]\right\}\nn\,.
\eeeq
The interference terms give
\beq\label{eq:interf12}
2\frac{g_s^4}{s_{12}s_{34}}\Bigl[
(g_Lg'_L+g_Rg'_R)(s_{14}^2+s_{23}^2)+
(g_Lg'_R+g_Rg'_L)(s_{13}^2+s_{24}^2)\Bigr]
\sum_{i<j} C_{ij} \frac{x_{ij}}{s_{i5}s_{j5}}\,,
\eeq
where $x_{12}=x_{34}=2s_{12}s_{34}$.  In the other cases there is a
contribution from $A^{(g)}_{hh',h''}$ and the results are slightly more
complicated:
\beeq
&&x_{13}=x_{24}=s_{14}s_{23}-s_{13}s_{24}-s_{12}s_{34}\,,\nn\\
&&x_{14}=x_{23}=s_{13}s_{24}-s_{14}s_{23}-s_{12}s_{34}\,.
\eeeq

\subsection{Results}

\FIGURE{
  \centering\centerline{
  \includegraphics[scale=0.8]{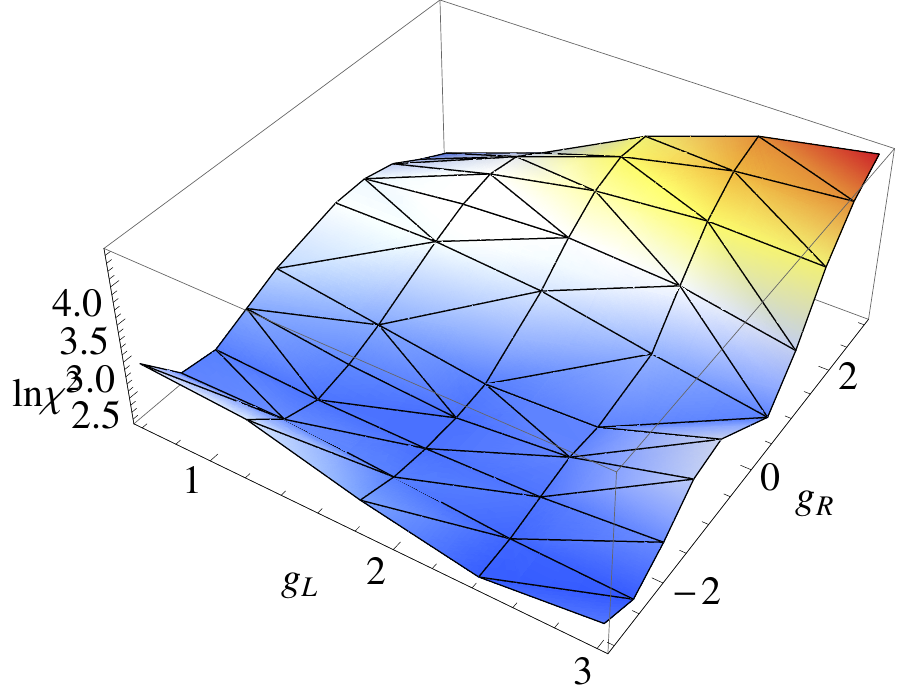}
  \includegraphics[scale=0.8]{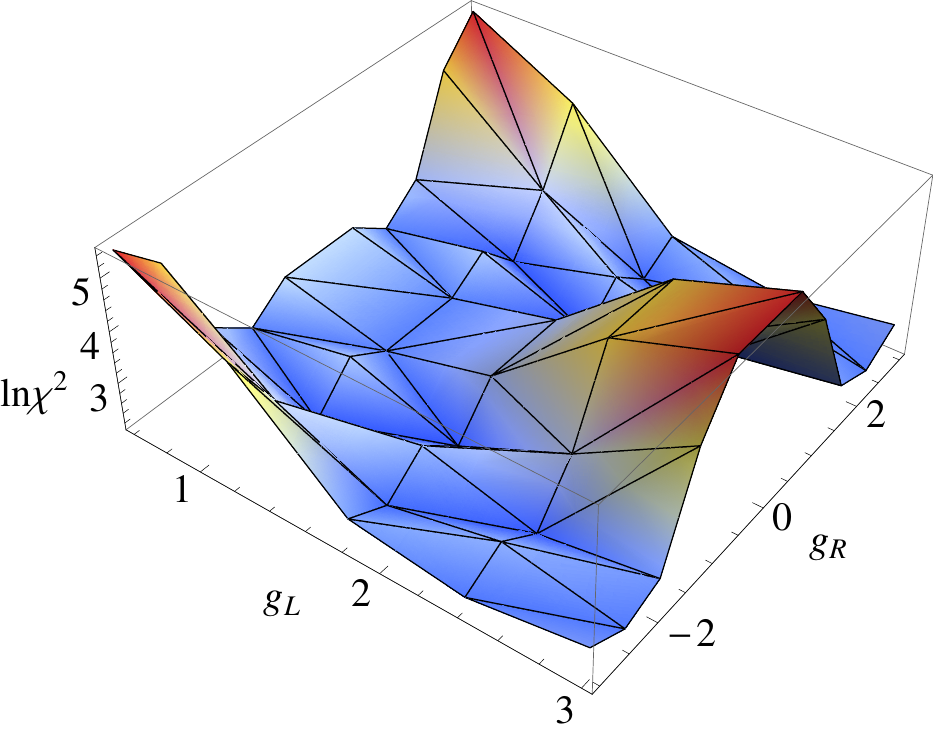}}
\vspace*{-5mm}
 \caption{\label{fig:scans0}%
As in Fig.~\protect{\ref{fig:scans}}, but using massless matrix
elements instead of the massive ones.}}

Figure~\ref{fig:scans0} shows the results of using the massless matrix
elements derived above instead of the massive ones in the fits to the
$t\bar t$ transverse momentum dependence of the asymmetry.  Here we
used the massive kinematics for each phase space point, and computed
the massless matrix elements from the actual massive $s_{ij}$ values
according to the expressions given.  Comparing with
Fig.~\ref{fig:scans}, we see that the essential qualitative features
of the predictions are captured by this simple prescription. 


\bibliography{QCDtasy.bib}
\bibliographystyle{utphys}

\end{document}